\documentclass[aps,twocolumn,english,superscriptaddress,citeautoscript,showkeys,preprintnumbers,amsmath,amssymb,floatfix,footinbib]{revtex4-2}
\usepackage{graphicx}
\usepackage{dcolumn}
\usepackage{amsmath}
\usepackage[version=4]{mhchem}
\usepackage{amsmath,array,graphicx}
\usepackage{array,multirow}
\usepackage{booktabs}
\usepackage{bm}
\DeclareGraphicsExtensions{.pdf,.png,.jpg}
\PassOptionsToPackage{export}{adjustbox}
\usepackage{mhchem}
\usepackage[utf8]{inputenc}
\usepackage{makecell}
\usepackage[colorlinks,allcolors=blue]{hyperref}
\usepackage{float}
\usepackage{setspace}
\usepackage{lipsum}
\usepackage{siunitx}
\usepackage{makecell}
\usepackage{csquotes}
\usepackage{gensymb}
\usepackage{array}
\usepackage{color}
\usepackage{placeins}
\usepackage{float}
\usepackage{graphicx}
\usepackage{subfigure}
\begin{document}
\title{Magnetic excitations and exchange parameters of a nickel chain compound PbMn$_2$Ni$_6$Te$_3$O$_{18}$: Neutron scattering and density functional theory studies}


\date{\today}
\author{S. Uthayakumar}
\email{Uthay.Sivaperumal@stfc.ac.uk}
\affiliation{ISIS Pulsed Neutron and Muon Source, STFC Rutherford Appleton Laboratory, Harwell Campus, Didcot, Oxon OX11 0QX, United Kingdom}
\affiliation{Department of Physics, Royal Holloway University of London, Egham, TW20 0EX, United Kingdom}

\author{D. T. Adroja}
\email{devashibhai.adroja@stfc.ac.uk}
\affiliation{ISIS Pulsed Neutron and Muon Source, STFC Rutherford Appleton Laboratory, Harwell Campus, Didcot, Oxon OX11 0QX, United Kingdom}
\affiliation{Highly Correlated Matter Research Group, Physics Department, University of Johannesburg, Auckland Park 2006, South Africa}

\author{Amit  Pokhriyal}
\affiliation{Raja Ramanna Centre for Advanced Technology, Indore - 452013, India}
\affiliation{Homi Bhaba National Institute, Anushaktinagar, Mumbai 400094, India.}
\author{A. K. Bera}
\affiliation{Solid State Physics Division, Bhabha Atomic Research Centre, Mumbai 40085, India} 
\affiliation{Homi Bhaba National Institute, Anushaktinagar, Mumbai 400094, India.}
\author{Haranath Ghosh}
\email{hngenator@gmail.com}
\affiliation{Raja Ramanna Centre for Advanced Technology, Indore - 452013, India}
\affiliation{Homi Bhaba National Institute, Anushaktinagar, Mumbai  400094, India.}

\author{Tatiana Gudi}
\affiliation{School of Science and Technology, Physics Division, University of Camerino, I-62032, Camerino, Italy}
\affiliation{ISIS Pulsed Neutron and Muon Source, STFC Rutherford Appleton Laboratory, Harwell Campus, Didcot, Oxon OX11 0QX, United Kingdom}
\author{Manh Duc Le}
\affiliation{ISIS Pulsed Neutron and Muon Source, STFC Rutherford Appleton Laboratory, Harwell Campus, Didcot, Oxon OX11 0QX, United Kingdom}

\author{Christian  Balz}
\affiliation{ISIS Pulsed Neutron and Muon Source, STFC Rutherford Appleton Laboratory, Harwell Campus, Didcot, Oxon OX11 0QX, United Kingdom}
\affiliation{Neutron Scattering Division, Oak Ridge National Laboratory, Oak Ridge, Tennessee 37831, USA}
\author{R. A. Ewings}
\affiliation{ISIS Pulsed Neutron and Muon Source, STFC Rutherford Appleton Laboratory, Harwell Campus, Didcot, Oxon OX11 0QX, United Kingdom}
\author{Minal Gupta}
\affiliation{Department of Physics, Indian Institute of Technology Indore,  Simrol, Khandwa road, Indore-453552, India}
\affiliation{Department of Mechanical Engineering, University of South Carolina, Columbia, South Carolina 29208, USA}

\author{P. R. Sagdeo}
\affiliation{Department of Physics, Indian Institute of Technology Indore,  Simrol, Khandwa road, Indore-453552, India}
\author{D. Prabhakaran}
\affiliation{Department of Physics, Clarendon Laboratory, Oxford University, Oxford OX1 3PU, United Kingdom}

\author{J.P. Goff}
\affiliation{Department of Physics, Royal Holloway University of London, Egham, TW20 0EX, United Kingdom}

\begin{abstract}
We have investigated the quasi-one dimensional Ni-chain compound PbMn$_2$Ni$_6$Te$_2$O$_{18}$ using theoretical DFT calculations, inelastic neutron scattering and optical spectroscopy in order to understand the nature of magnetic exchange interactions. Our inelastic neutron scattering study at 5 K on a powder sample reveals two bands of magnetic excitations, the first near 8 meV and the second near 18 meV originating from the antiferromagnetic zone center near $Q$ = 1~\AA. On the other hand at 100 K (which is above T$_N$ = 86 K) a broad diffuse scattering signal is observed indicating the presence of short range magnetic correlations. We have analyzed the magnetic excitations based on the Linear Spin Wave Theory (LSWT) and compared the experimentally estimated exchange parameters with the DFT calculations. Our analysis reveals that the value of the exchange parameter at the larger distance (d=3.654 $\AA$) $J_3$=4.21(8) meV between Ni-Ni (from inter-chain) is the strongest amongst the allowed six exchange parameters, which suggests that this system is not really a quasi-one-dimensional and confirmed by the absence of a Haldane gap. We have also presented the electronic structure calculations. The spin-polarized partial density of states (DOS) projected onto the Mn-d and Ni-d  orbitals reveals that the Ni-d$_{x^2-y^2}$ contribution is dominant below the Fermi level in the spin-up and spin-down channel, while a minimal contribution from spin-up Mn states in the occupied region, suggesting a nearly high-spin state. The estimated Néel temperature, based on experimental exchange parameters is found to be in close agreement with the experimental value. 

\end{abstract}
\maketitle
\section{INTRODUCTION}
\label{Intro}
 The search for realizations of quantum spin systems where magnetic exchange between localized spins is restricted to one (1D) or two (2D) spatial dimensions is important for the verification of modern theories of quantum magnetism and for the exploration of novel magnetic phenomena and topological behavior ~\cite{2004LNP6451M, Sutherland, PhysRevA.75.032340}. They display quantum phase transitions (QPTs) and quantum critical behavior at extremely low temperatures ~\cite{Sachdev2011, PhysRevLett.90.186601}. Therefore, low-dimensional (LD) magnetic materials have attracted a great deal of interest recently, because of their anomalous magnetic properties at low temperatures ~\cite{PhysRevB.107.224412, Steiner01031976, RevModPhys.84.1253, Greiter2002}. Among LD systems, the subject of one-dimensional (1D) spin-chain systems has attracted considerable attention due to the well-known fact that a  uniform half-integer spin chain with an isotropic Heisenberg exchange interaction does not undergo long-range magnetic order (LRO) at finite temperatures due to strong quantum spin fluctuations ~\cite{PhysRevLett.17.1133, IAffleck_1989, GIAMARCHI}. Quantum spin fluctuations in a 1D or frustrated magnet destabilize the magnetic ground state, while keeping sizable spin correlations, and hence results in a new ground state called a quantum spin liquid (QSL) which is stabilized for low spin values~\cite{PhysRevB.12.3908, PhysRevLett.45.1358, PhysRevLett.133.236702, SubirSachdev}.  The excitation spectrum of a half-integer valued spin chain is gapless ~\cite{PhysRevLett.45.1358, IAffleck_1989}. In a one-dimensional spin-1/2 chain, particle-like excitations are observed, which are called \emph{spinon}~\cite{PhysRevB.101.235107, PhysRevB.96.035127, PhysRevB.96.054423}, responsible for spin fluctuations in the paramagnetic state. For an antiferromagnetic Heisenberg
model in one dimension, in 1931 Bethe introduced his famous “ansatz” method to find the exact quantum ground state ~\cite{Bethe1931}.  The mechanism of gapless excitations was given by the Lieb-Shultz-Mattis theorem whereby the separation between the ground and first excited state energies of a half-integer spin chain was shown to vanish in the thermodynamic limit ~\cite{LIEB1961407}. For anisotropic exchange in the Ising limit the spin chain achieves an ordered ground state at zero temperature ~\cite{PhysRevX.7.031061}.

On the other hand for a one-dimensional (1D) Heisenberg spin chain with integer spins and antiferromagnetic (AFM) nearest neighbor coupling Haldane predicted that the ground state of such a system would be a nonmagnetic singlet state which would be separated in energy from the excited triplet state by a gap (E$_g$) ~\cite{HALDANE1983464,PhysRevLett.50.1153,IAffleck_1989}, which was later confirmed through the experimental studies ~\cite{PhysRevLett.56.371, PhysRevB.38.543}. Another incredibly special feature of the Haldane gap state is the existence of long-range correlations that lead to edge modes, which have been observed in experiments ~\cite{JOLICOEUR2019445, PhysRevLett.65.3181}. The magnetic susceptibility approaches zero at zero temperature in Haldane gap systems, while the magnetic susceptibility decreases to finite values in the case of half-integer spin. In practice, it is difficult to realize ideal 1D Haldane gap systems as often the inter-chain interactions will be of comparable strength to the intra-chain interactions in many systems, resulting in a long range magnetic ground state. Since Haldane's conjecture, many theoretical and experimental studies have been carried out to verify it. From Monte Carlo calculations the relation between the Haldane gap (E$_g$) and exchange parameter ($J$) is predicted to be E$_g$=0.41/J~\cite{U.Schollwöck_1995} for $S=1$. The same calculations predict the ratio (E$_g$/J) for an $S=2$ Haldane chain to be very small~\cite{U.Schollwöck_1995}. 

The ideal systems with an $S=1$ Haldane gap are based on Ni$^{2+}$ ($3d^8$, $S=1$) ions~\cite{Tin2023, PhysRevResearch.2.013082, JeanPierreRenard} but some are also found with Ru$^{4+}$ ~\cite{Tl2Ru2O7}. Experimental evidence for Haldane gaps has been obtained in CsNiC1$_3$ ~\cite{PhysRevLett.56.371,PhysRevB.38.543},  Y$_2$BaNiO$_5$ ~\cite{PhysRevLett.73.1857, DARRIET1993409}, Ni(C$_2$H$_8$N$_2$)$_2$NO$_2$ClO$_4$ ~\cite{J.P.Renard_1987}, SrNi$_2$V$_2$O$_8$ ~\cite{Bera_2012, Bera_2013, Bera_2015, Bera_2015a}, and Tl$_2$Ru$_2$O$_7$ ~\cite{Tl2Ru2O7}. In CsNiC1$_3$ antiferromagnetic (AF) long-range order below T$_N$ = 4.9 K has also been found as a result of weak inter-chain coupling ~\cite{PhysRevLett.56.371, PhysRevB.38.543}. In SrNi$_2$V$_2$O$_8$, inter-chain couplings reduce the energy gap and bring the system close to the phase boundary between gapped and gap-less states ~\cite{Bera_2015}. The Haldane gap and magnetic ground states were also reported in the rare earth and Ni-based spin-chain, RE$_2$BaNiO$_5$ (RE is rare earth, Pr Yb to Dy)~\cite{PhysRevLett.73.1857,PhysRevB.54.6437}.

 The most investigated Haldane spin chains have isolated $S=1$ chains of Ni$^{2+}$ or Ru$^{4+}$ ions and no other magnetic transition metal ions (i.e. mixed $3d$-ions). Thus the investigations of Haldane spin chains (Ni$^{2+}$ $S=1$) along with different $3d$ magnetic ions in a quasi-one dimensional arrangement could provide an opportunity to discover an exotic magnetic ground state. Recently, the crystal structure and magnetic properties of such a two-spin chain (with $S=3/2$ of Mn$^{2+}$ and $S=1$ of Ni$^{2+}$) in PbMn$_2$Ni$_6$Te$_3$O$_{18}$ have been  reported ~\cite{WEDEL19992193, acs.inorgchem.pbmn2}. PbMn$_2$Ni$_6$Te$_3$O$_{18}$ crystallizes in the hexagonal crystal system, space group, P6$_3$/m (No. 176), and consists of NiO$_6$ and TeO$_6$ octahedra forming a double chain structure running along the c-axis Fig.~\ref{FIG:Unitcell}(a-b)~\cite{WEDEL19992193, acs.inorgchem.pbmn2}. 
 
 On the other hand  Mn$^{2+}$ ions are trigonal-prismatic coordinated by O$^{2-}$ and located in the trigonal tunnel built by three Ni chains, which results in a complex exchange network. Interestingly, in the hexagonal $ab$ plane the chains have triangular symmetry, which may produce geometrical magnetic frustration in the inter-chain magnetic interactions pathways as shown in Fig.~\ref{FIG:Unitcell}(b). Magnetic susceptibility and heat capacity measurements revealed a long-range magnetic ordering transition at $T_N$ = 86~K ~\cite{acs.inorgchem.pbmn2}. Neutron diffraction measurements revealed a collinear antiferromagnetic arrangement of Mn$^{2+}$ ions in the Mn$_2$O$_9$ dimeric unit and Ni$^{2+}$ ions in the zigzag chain ~\cite{acs.inorgchem.pbmn2}. The ordered magnetic moments (along the $c$-axis), $\mu_{Mn}$ = 4.71(3)$\mu_B$ and $\mu_{Ni}$ = 1.96(2) $\mu_B$, indicate that both magnetic moments are fully ordered below the ordering temperature. A Raman spectroscopy study on this compound reveals an anharmonic phonon behavior and a coupling to the spin system at $T_N$ = 84 K, as well as weak anomalies at T=200 K ~\cite{PhysRevB.103.064401}. Furthermore, from the magnetic Raman scattering observed at 240 cm$^{-1}$  an exchange coupling of $J$ = –86 K was estimated, which is in agreement with the Curie-Weiss temperature. Very recently, the linear magneto-electric effect in the c-axis collinear antiferromagnetic phase and spontaneous nonreciprocal directional dichroism for unpolarized light propagating along the $c$-axis  have been observed, suggesting a magnetic toroidal moment along the $c$-axis in PbMn$_2$Ni$_6$Te$_3$O$_{18}$ ~\cite{JPhysSocJpn93}.

 It would therefore certainly be a fruitful playground for further exploration of this system using microscopic measurements, such as inelastic neutron scattering (INS) combined with DFT theoretical calculations. The motivation was to test the prediction of Haldane gap value for integer spin chain  (Ni$^{2+}$ S=1) in presence of the exchange interactions between non-integer spins (Mn$^{2+}$ S=5/2, Mn-Mn spin-dimers). Further, motivation was to test the exchange parameters predicted from the DFT calculations using the measured excitations from the inelastic neutron scattering as well to compare the band gap value from the DFT theory with that obtained from the optical spectroscopy. Here we represent our experimental study of INS, optical spectroscopy and DFT calculations on PbMn$_2$Ni$_6$Te$_3$O$_{18}$, which reveal that the larger distance (d=3.758 $\AA$) Ni-Ni exchange interactions are the strongest among the allowed six exchange interactions.

\section{EXPERIMENTAL METHODS}
\label{expt}
The synthesis of PbMn$_2$Ni$_6$Te$_3$O$_{18}$ was carried out using a standard solid state reaction process. In this process, high-purity (99.99\%) chemicals of PbO, MnO, NiO and TeO$_2$ were mixed in a stoichiometric ratio with a nominal composition of PbMn$_2$Ni$_6$Te$_3$O$_{18}$. The mixture was finely powdered in an agate mortar and calcined in air at 650–750 $^{\circ}$C for 30 h with intermittent grinding to ensure chemical homogeneity. The calcined powders were then sealed in an evacuated quartz ampoule to reduce volatilization. Subsequently, the ampoule was heated in the temperature range of 750-800 $^{\circ}$C for 15 h to yield a single phase of polycrystalline PbMn$_2$Ni$_6$Te$_3$O$_{18}$. The single-phase nature of the sample was checked using Bruker D8 Discover X-ray Diffraction System using Cu K$_\alpha$ radiation. The magnetic susceptibility was measured using a Quantum Design Magnetic Properties Measurement System (MPMS).

The inelastic neutron scattering (INS) measurements were performed on the neutron time-of-flight spectrometers LET ~\cite{BEWLEY2011128} and  MARI ~\cite{LE2023168646} at the ISIS facility.  On both MARI and LET the INS spectra were recorded at temperatures of 5 and 100 K. Experiments on both instruments were done with a powder sample of mass ~ 3.5 g. The powder sample was placed in an annular aluminum sample can using a thin Al-foil envelope and then cooled down to 5 K using a closed-cycle refrigerator in He-exchange gas for the low temperature measurements on both MARI and LET.  On MARI data were collected with incident neutron energies (E$_i$) of 60 and 17.7 meV and a Gd-Fermi chopper at 200 Hz, having elastic resolution $\Delta$E=2.52 meV and 0.50 meV, respectively using repetition-rate multiplication (RRM) mode. These results are presented in the Appendix-B. On LET we used primary E$_i$= 22.8 meV ($\Delta$E=1.06 meV), with the disc choppers set at 250 Hz and 120 Hz, which also gave data of E$_i$=7.5, 3.7 and 2.2 meV (with $\Delta$E=0.21 meV, 0.08 meV and 0.04 meV, respectively) in the same run, each of the INS patterns were measured for 6 h (1000 $\mu$A of integrated proton current). The large detector banks of the MARI and LET spectrometers allow a simultaneous measurement over a large area in momentum (Q) and energy ($\omega$) space. The INS data were reduced using the {\tt Mantid} software package~\cite{ARNOLD2014156}. The raw data were corrected for detector efficiency and time-independent background following standard procedures. 

The spin-wave simulations were carried out using the {\tt SpinW} program's  implementation of linear spin wave theory ~\cite{Toth_2015}. The exchange parameters were determined by fitting the 2D experimental spectra. The simulated spectra were convoluted using the instrument resolution estimated from the {\tt PyChop} program in {\tt Mantid}~\cite{ARNOLD2014156}.

\section{Computational methods}
The first principles Density Functional Theory (DFT) calculations were performed using plane wave pseudo potential method implemented in the Cambridge Serial Total Energy Package (CASTEP)~\cite{ClarkSegallPickardHasnipProbertRefsonPayne+2005+567+570}. The electronic exchange correlation functional in the generalized gradient approximation (GGA) ~\cite{PhysRevLett.77.3865}, parameterized according to Perdew–Burke–Enzerhof (PBE) ~\cite{10.1063/1.472933}, was employed. Electron correlation corresponding to a $3d$ transition metal ion is not well described by DFT calculations and may predict metallic electronic structure for a magnetic insulating system. This inadequacy of DFT is empirically corrected either by adding on-site Coulomb repulsion U on the transition metal site \cite{dudarev1998} or by adding a Hartree-Fock exchange potential. In the PbMn$_2$Ni$_6$Te$_3$O$_{18}$ compound, the experimental band-gap value (see, optical study in~ Appendix-A) was used as a reference to tune the Hubbard $U$ values for the two magnetic ions, Ni$^{2+}$ and Mn$^{2+}$. A range of 3-7\,eV was scanned for the Hubbard $U$ values relative to the band gap for the magnetic centers Ni$^{2+}$ and Mn$^{2+}$. A Hubbard $U$ value of 7\,eV for Ni and 4\,eV for Mn was used in all calculations, which are close to 6.2\,eV and 3.9\,eV, respectively reported in the literature~\cite{HubbardU}. The calculated band gap in the relaxed lowest-energy magnetic configuration with these Hubbard $U$ values was found to be 1.52\,eV, which is close to the experimental band gap of 1.82-1.91\,eV (see Appendix-A). In order to calculate the six superexchange parameters (J$_1$-J$_6$), (Local Spin Density Approximation) LSDA+U calculations for seven different spin configurations were performed. A kinetic energy cutoff of 600 eV was kept for the plane-wave expansion after convergence testing. Brillouin zone integration was performed with a 4x4x4 Monkhorst-Pack k-point mesh. The magnetic symmetry of all seven configurations were determined from the individual spin of each magnetic ion. The spin was flipped at each individual magnetic center to get one FM and six different AFM configurations. 

\section{Calculations of Exchange parameters and N\'eel Temperature}
\begin{figure}
	\centering 
\includegraphics[width=\linewidth, trim= 0 0  40 0]{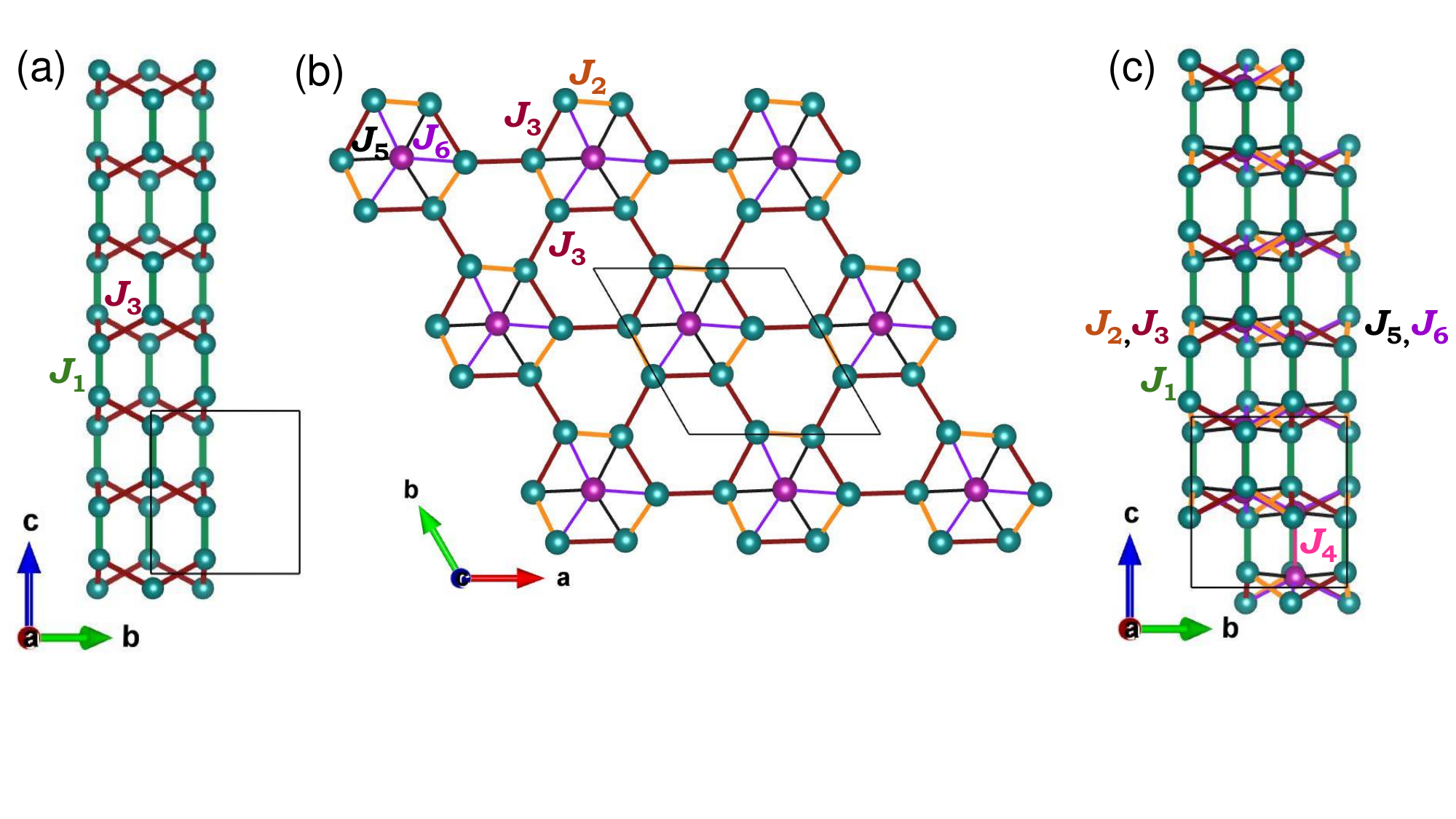}
	\caption{(Color online) (a) The Ni$^{2+}$ spin-chains in  PbMn$_2$Ni$_6$Te$_3$O$_{18}$ along the $c$-axis, considering the strongest J$_1$ and J$_3$ exchange interactions (For details see Table ~\ref{I}). (b) The exchange interaction topology in the ab plane (Ni$^{2+}$ light green sphere and   Mn$^{2+}$ purple sphere), involving J$_2$, J$_3$, J$_5$, and J$_6$ exchange interactions, which forms a maple leaf lattice ~\cite{PhysRevB.98.064412}. (c) The exchange interaction topology considering all the six exchange interactions.}
	\label{FIG:Unitcell}
\end{figure}

 The six superexchange interaction pathways we consider in PbMn$_2$Ni$_6$Te$_3$O$_{18}$ unit cell and their corresponding bond angle/distance is shown in table \ref{I} and Fig.~\ref{FIG:Unitcell}. Calculations were performed in the fully relaxed structure, where in order to minimize atomic forces, all the atomic positions and cell parameters were set free to move during optimization. 
\begin{table*}
	\caption{\bf{Exchange interaction pathways, bond lengths, bond angles, exchange parameters and calculated Néel temperatures for PbMn$_2$Ni$_6$Te$_3$O$_{18}$.}}
	\centering 
	\begingroup
	\renewcommand{\arraystretch}{1.4}
	\setlength{\tabcolsep}{5pt}                     
	\begin{tabular}{l c c c c c l }		 
		\hline    
		\bf{}   &\bf{interaction} & \bf{Bond length} & \bf{bond angle} & \bf{DFT-exchange} & \bf{Exp. Exchange}\\
		& \bf{pathway} & \bf{(\AA)}  & \bf{(deg)} & \bf{(meV)} & \bf{(meV)}\\\hline 
		\endgroup
		\begingroup
		\renewcommand{\arraystretch}{1}{J$_1$}  & Ni-O$_3$-Ni & 2.943   & 84.378 &  \\
		{}  & Ni-O$_4$-Ni & 2.943   & 91.338 &  5.423  & 2.618 (58)&  \\
		{J$_2$}  & Ni-O$_2$-Ni & 3.238 &  99.429 & 3.125 & 1.23 (12)& \\
		{J$_3$}  & Ni-O$_1$-Ni & 3.758 &  123.673 & 3.980 & 4.212 (78)& \\
		{J$_4$}  & Mn-O$_3$-Mn & 3.424 & 91.887 & 2.788 & 2.120 (14) &\\
		{J$_5$}  & Mn-O$_2$-Ni & 3.027  & 89.224 &   \\
		{} & Mn-O$_3$-Ni & 3.027  & 82.769 & -0.721 &-1.096 (19) \\
		{J$_6$} & Mn-O$_2$-Ni & 3.469  & 106.345 &  1.946 & 0.902 (88) \\
        \hline
		& & & & \multicolumn{2}{l}{\bf{Néel Temperature (in K), using}}  \\
         & & & & DFT-exchange & Exp.-exchange \\
		\hline
		\multicolumn{2}{l}{} S$_{Mn}$(S$_{Ni}$) = 2.295(0.855) (Hirshfeld analysis \cite{Hirshfeld1977}) & & & 119.92 & 89.24 $\pm$ 10.03  \\
		\multicolumn{2}{l}{} S$_{Mn}$(S$_{Ni}$) = 2.465(0.855) (Mulliken anlaysis \cite{PhysRevB.54.16317}) & & & 131.59  & 97.87 $\pm$ 10.55  \\
		\multicolumn{1}{l}{}S$_{Mn}$(S$_{Ni}$) = 5/2(1) & & &
								& 149.00 & 110.94 $\pm$ 13.08 \\
		\hline
	\end{tabular}
	\endgroup
	\label{I}  
\end{table*}

For each magnetic configuration, we calculate the magnetic energy according to the Heisenberg Hamiltonian,

\begin{equation*}
H = \sum_{ij}J_{ij}S_iS_j,
\end{equation*}  

\noindent where J$_{ij}$ is the spin exchange parameter for the superexchange interaction between the spin sites $i$ and $j$, while $S_i$
and $S_j$ are the spin angular momentum operators at the $i^{th}$ and $j^{th}$ magnetic centers.
By mapping the total energy differences between the ordered spin configurations obtained from the DFT calculations on to the corresponding energy differences obtained from the Heisenberg Hamiltonian, we can estimate the values of exchange parameters \cite{BaTh2,magnetochemistry9070164}. For the six exchange interactions J$_1$, J$_2$, ..., J$_6$; we must consider seven ordered spin configurations with energies,

\begin{align}
\text{FM} &= 6J_1 + 6J_2 + 12J_3 + \frac{25}{2}J_4 + 30J_5 + 30J_6 + \text{NM}, \label{eq:FM} \\
\text{AFM1} &= 6J_1 + 6J_2 + 12J_3 + \frac{25}{2}J_4 - 30J_5 - 30J_6 + \text{NM}, \label{eq:AFM2} \\
\text{AFM2} &= 6J_1 + 6J_2 - 12J_3 + \frac{25}{2}J_4 - 10J_5 - 10J_6 + \text{NM}, \label{eq:AFM3} \\
\text{AFM3} &= -6J_1 - 6J_2 - 12J_3 - \frac{25}{2}J_4 + 30J_5 - 30J_6 + \text{NM}, \label{eq:AFM1} \\
\text{AFM4} &= -6J_1 - 6J_2 - 12J_3 + \frac{25}{2}J_4 + \text{NM}, \label{eq:AFM4} \\
\text{AFM5} &= -6J_1 - 6J_2 + 12J_3 - \frac{25}{2}J_4 + 10J_5 + 30J_6 + \text{NM}, \label{eq:AFM5} \\
\text{AFM6} &= 6J_1 - 6J_2 - 12J_3 + \frac{25}{2}J_4 - 30J_5 + 30J_6 + \text{NM}. \label{eq:AFM6}
\end{align}

\noindent where FM(ferromagnetic), AFM1 (anti-ferromagnetic 1), ..., AFM6, are energies of different magnetic configurations (see also Fig.\ref{FIG:Spin_config} in the Appendix-E) and NM stands for energy corresponding to non-magnetic unit cell. Solving these linear equations, we obtain J$_1$ = 5.423 meV, J$_2$ = 3.125 meV, J$_3$ = 3.980 meV, J$_4$ = 2.788 meV, J$_5$ = -0.721 meV, J$_6$ = 1.946 meV. Positive J corresponds to antiferromagnetic (AFM) interaction and negative J to ferromagnetic (FM) interaction. It is evident that all the J values support the experimental findings, indicating that the magnetic moments of both Mn$^{2+}$ and Ni$^{2+}$ align antiferromagnetically \cite{acs.inorgchem.pbmn2}. 

The FM sign of J$_5$ is in full agreement with the magnetic structure. The DFT overestimates the magnitude of the exchange parameters, most likely because of inherent approximations in the exchange-correlation functional used in the calculation  \cite{koo_crvo4,KOO2009602,grau2003cation}. The J$_5$ superexchange interaction occurs between the half-filled Mn$^{2+}$ ions and the more than half-filled Ni$^{2+}$ ions. One of the angles, Mn-O$_2$-Ni, is close to 90 degrees, favoring a ferromagnetic interaction. The strongest exchange estimated from the DFT is AFM between Ni-O$_3$-Ni and Ni-O$_4$-Ni at the shortest bond length of 2.943~$\AA$, while the second strongest exchange for the largest bond length of 3.758~$\AA$~for Ni-O$_1$-Ni.

Additionally, the Néel temperature ($T_N$) is presented in Table~\ref{I}, calculated within the mean-field approximation \cite{PhysRevB.80.014408} using both DFT-derived and experimentally estimated spin-wave exchange parameters. The values in the first two rows correspond to cases where the magnetic moment per magnetic center is determined from DFT outputs using Hirshfeld ~\cite{Hirshfeld1977} and Mulliken ~\cite{PhysRevB.54.16317} population analysis, respectively. The estimated Néel temperature, based on experimental exchange parameters and the magnetic moment obtained from Hirshfeld population analysis, is found to be in close agreement with the experimental value of 86~K.

\begin{figure*}
	\includegraphics[width=\linewidth]{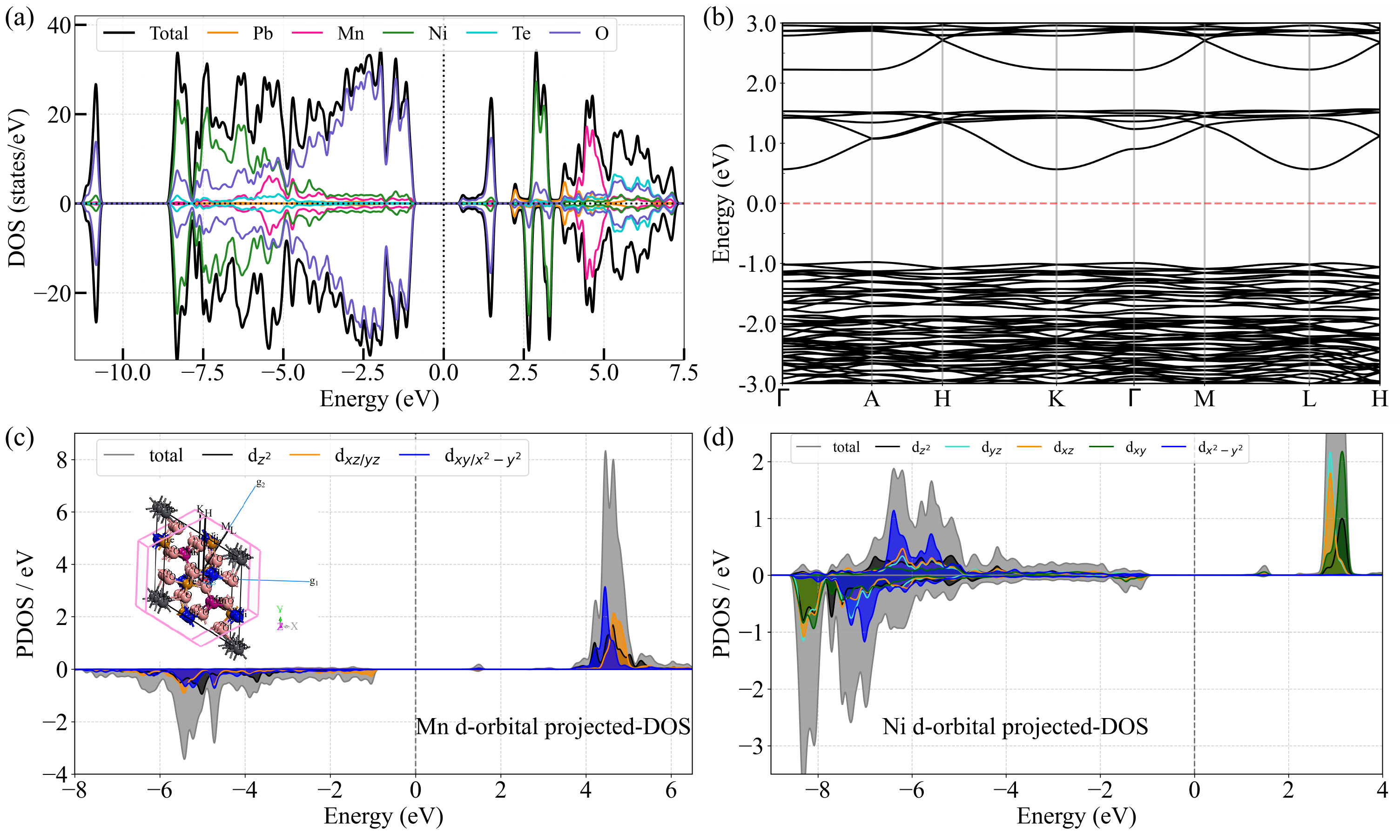}
	\caption{(Color online) Electronic structure calculations of PbMn$_2$Ni$_6$Te$_3$O$_{18}$. (a) The total density of state (DOS) of Pb, Mn, Ni, Te and O, (b) The electronic band structure using GGA+U. (c) Mn-d band partial projected DOS, (d) Ni-d band partial projected DOS.  The inset in (c) shows the high symmetry points in the Brillouin Zone (BZ).}
	\label{BS}
\end{figure*} 

\section{Electronic structure}
Figure~\ref{BS}(a) shows the atom-projected density of states for the \ce{PbMn_2Ni_6Te_3O_{18}} compound in the antiferromagnetically coupled Ni chain configuration, which has the lowest energy among the considered configurations. The energy gap found around the Fermi level indicates the insulating nature, which is in agreement with the optical study (see Appendix-A). The low-energy occupied states are primarily dominated by contributions from oxygen and nickel, with O states being more prominent near the Fermi level (from -5 to -1 eV) and Ni states dominating the higher binding energy range (from -10 to -6 eV). Notably, manganese states also contribute significantly within the -5 to -1 eV energy interval. In the unoccupied energy region, the states are successively dominated by oxygen, nickel, manganese, and tellurium. As the system is in the antiferromagnetic ground state, the contribution due to each element is same in the two spin-channels (spin-up and spin-down).

Figure~\ref{BS}(b) illustrates the band structure along the high-symmetry points, which are depicted in the Brillouin zone in the inset of Figure~\ref{BS}(c). It is evident that most of the bands are relatively flat, while the band just above the energy gap exhibits a slightly higher degree of dispersion.

Figures~\ref{BS}(c) and \ref{BS}(d) display the spin-polarized partial density of states (DOS) projected onto the Mn-d and Ni-d orbitals, respectively. For the Ni orbitals, the Ni-d$_{x^2-y^2}$ contribution is dominant below the Fermi level in the spin-up and spin-down channel. In contrast, as shown in Figure~\ref{BS}(c), a minimal contribution from spin-up Mn states in the occupied region, suggesting a nearly high-spin state.

\section{EXPERIMENTAL RESULTS AND DISCUSSION}
\begin{figure} 
\centering
\includegraphics[width = \linewidth, trim= 50 00 80 70]{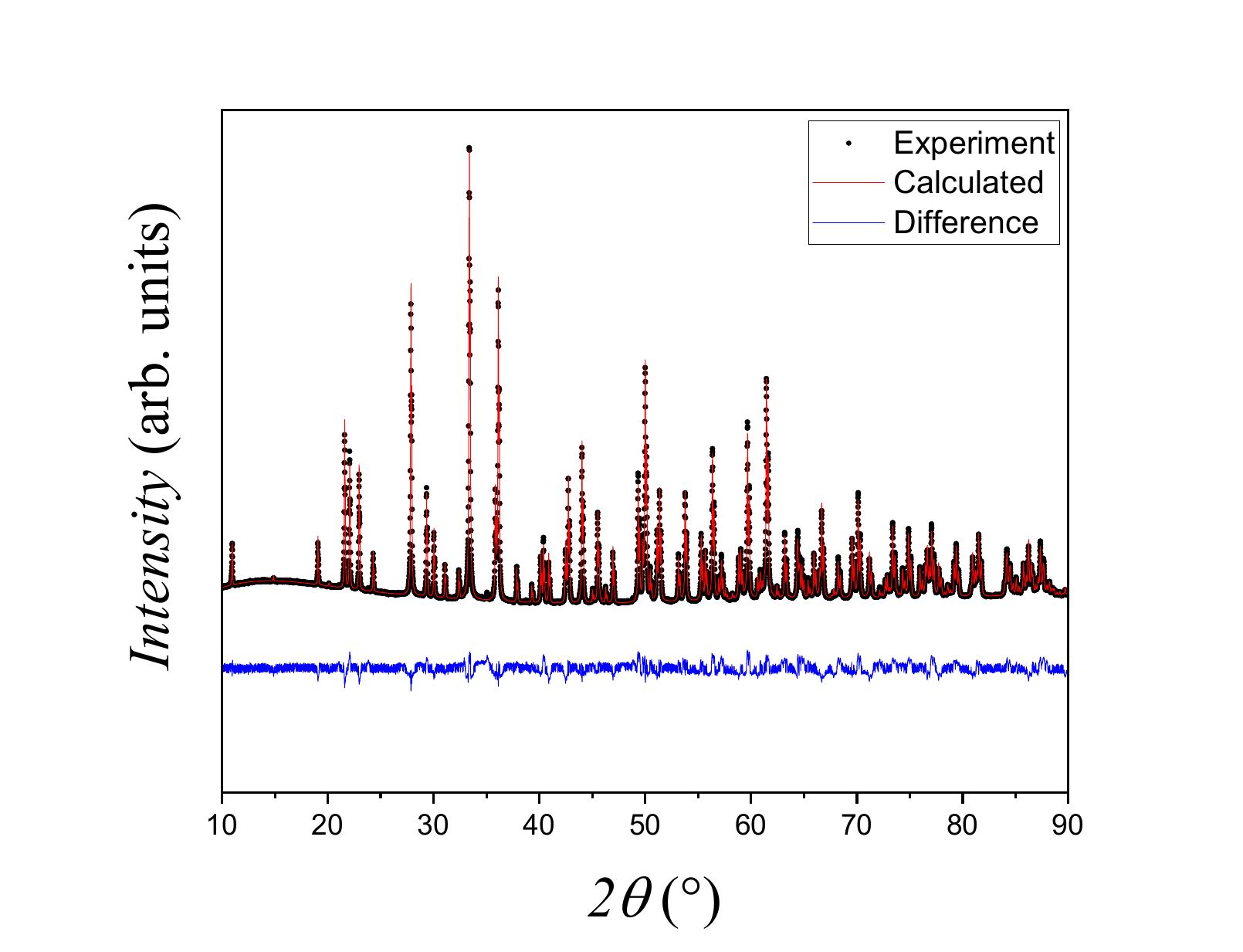}
\caption{(Color online)  Rietveld refined powder x-ray diffraction pattern for PbMn$_{2}$Ni$_6$Te$_2$O$_{18}$ ~at 300~K, along with the difference (Y$_{\mathrm{obs}}$-Y$_{\mathrm{cal}}$, shown by the blue solid  line at bottom).
}
\label{FIG:PXRD}
\end{figure}

\begin{figure} 
\centering
\includegraphics[width = \linewidth, trim= 50 80 80 70]{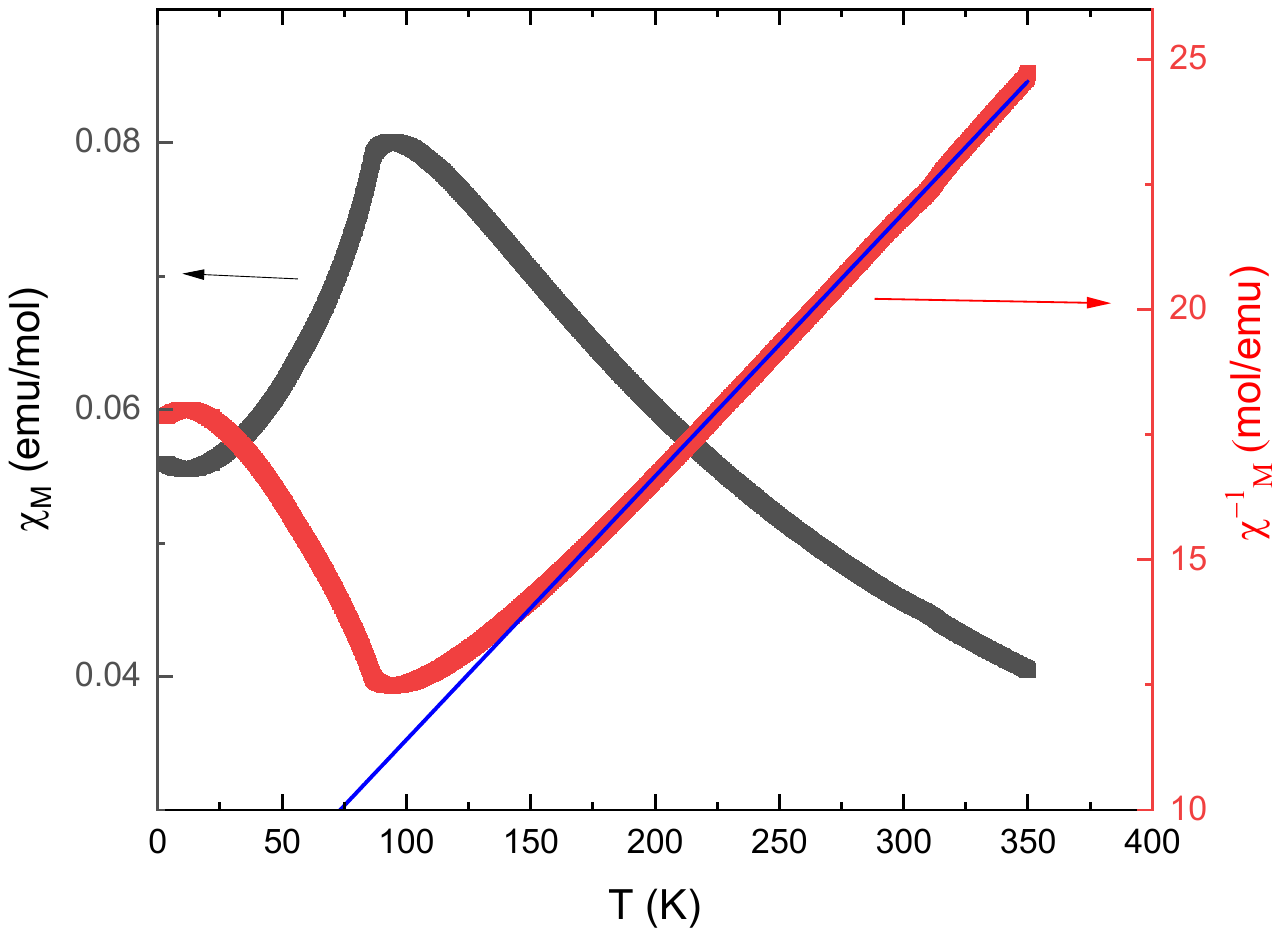}
\caption{(Color online)  (left) Magnetic susceptibility versus Temperature (right) Inverse magnetic susceptibility versus temperature for PbMn$_{2}$Ni$_6$Te$_2$O$_{18}$. The solid line shows the Curie-Weiss fit to the data.}
\label{FIG:SUS}
\end{figure}

The Rietveld refined powder x-ray diffraction (PXRD) pattern of PbMn$_{2}$Ni$_6$Te$_2$O$_{18}$ at 300~K (see Fig.~\ref{FIG:PXRD}) confirms the single phase nature of the sample, with the $P6_{3}/m$ (No 176) space group in agreement with an earlier report by Doi {\it et al.}~\cite{acs.inorgchem.pbmn2}. The refined lattice  parameters a=9.2863(3) $\AA$, and c=8.8180(2) $\AA$ are in good agreement with values reported in the literature ~\cite{acs.inorgchem.pbmn2}.

Figure~\ref{FIG:SUS} shows the temperature dependence of the magnetic susceptibility (left y-axis) and inverse magnetic susceptibility (right y-axis). The susceptibility exhibits a broad peak near 92~K indicating the antiferromagnetic ordering of Ni and Mn moments, which is in agreement with previous studies ~\cite{acs.inorgchem.pbmn2}. The inverse susceptibility exhibits a Curie-Weiss behavior between 150 K and 350 K and exhibits deviation from it below 150 K due to short range magnetic correlations. The value of effective magnetic moment estimated from the Curie-Weiss constant is $\mu$$_{eff}$=11.96 (2) $\mu$$_B$/f.u. and the Curie-Weiss temperature $\theta$$_P$= -100.5 (3) K. The negative value of $\theta$$_P$ indicates the presence of strong AFM interactions. Assuming the spin-only moment on Mn$^{2+}$ with $S=5/2$ ($\mu$$_{Mn^{2+}}$=5.92 $\mu$$_B$), we estimated the effective moment on Ni$^{2+}$ to be $\mu$$_{Ni^{2+}}$=3.33 (2)  $\mu$$_B$, which is slightly larger than the spin-only value of 2.83 $\mu$$_B$ for Ni$^{2+}$ (d$^8$). The larger value of $\mu$$_{Ni^{2+}}$=3.28 $\mu$$_B$ was also reported in PbCd$_{2}$Ni$_6$Te$_2$O$_{18}$ where the moment is only on Ni-ion ~\cite{acs.inorgchem.pbmn2}, which could be attributed to effect of spin-orbital coupling.

\begin{figure*}[!htb]
	\centering
	\includegraphics[width = \linewidth, trim= 0 130 0 50]{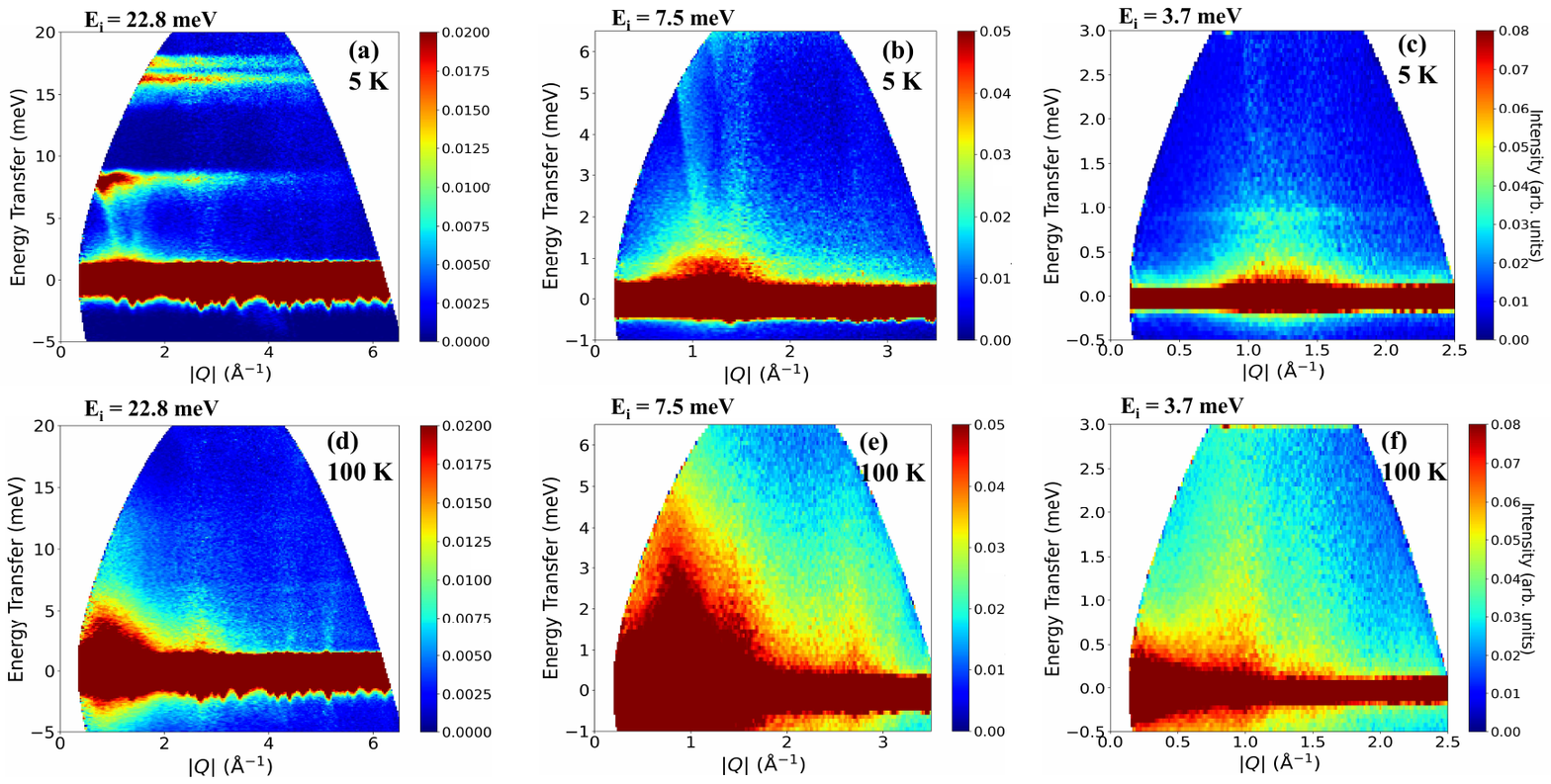}
	\caption{(Color online) Color coded inelastic neutron scattering intensity maps, energy transfer vs momentum transfer ($Q$) (a-c) at 5~K and (d-e) at 100 K of PbMn$_{2}$Ni$_6$Te$_2$O$_{18}$ measured with an incident energies $E{_i}$=22.8~meV, 7.5~meV and 3.7~meV, respectively  on LET.}
	\label{FIG:ins}
\end{figure*}	

The results of INS measurements to investigate the magnetic excitations of PbMn$_{2}$Ni$_6$Te$_2$O$_{18}$  below $T_\textrm{N}$ are depicted in Fig.~\ref{FIG:ins} (a-f). At $T = 5$~K,  we have seen well defined spin wave excitations. The first dispersive band is near 8~meV, while there are two non dispersive bands of excitations near 16 and 18~meV. The low incident energy measurements reveal two interesting features: firstly, as shown in Fig.~\ref{FIG:ins}(b), the low energy spin wave exhibits a linearly dispersing excitation originating from the AFM zone center and, secondly the presence of diffusive magnetic excitations centered near Q=1.3 \AA$^{-1}$ at 5 K, see Fig.~\ref{FIG:ins}(b-c). There is also a weak non-dispersive band near 1~meV (Fig.~\ref{FIG:ins}(c)), the origin of which will be discussed later. We were able to resolve two bands of excitations near 18 meV. 

Comparing the elastic Q-cuts at 5 K and 100 K (not shown here), we note that additional magnetic Bragg peaks develop, with the first at $|Q| = 1.055$~\AA$^{-1}$ (d=5.953 \AA) ~below $T_\textrm{N}$, corresponding to the (1 0 1) magnetic Bragg peak, which is consistent with the magnetic structure previously reported, with the propagation vector $\bm{k}=(0~0~0)$ ~\cite{acs.inorgchem.pbmn2}. 

The observed spin wave excitations seem to be gapless at least within the instrumental resolution, see Fig.~\ref{FIG:ins}(c). We also have data with E$_{i}$ = 2.2~meV (see the Appendix-C) with elastic resolution $\Delta$=0.032~meV, which also did not reveal any clear sign of a gap in the spin wave excitations at 5 K.  Furthermore, the E$_{i}$ = 2.2~meV data at 5~K reveal the presence of a flat excitation at 1 meV and diffuse scattering similar to that observed in E$_{i}$ = 3.7~meV data shown in Fig.~\ref{FIG:ins}(c)). This confirms that the 1~meV mode is real and not an artifact. As the temperature is increased to 100~K (i.e. above the $T_\textrm{N}$), we observed the presence of strong magnetic diffuse scattering due to the presence of short range magnetic correlations and no clear sign of spin waves. 

 \begin{figure*}[!htb]
	\centering
	\includegraphics[width = \linewidth, trim= 0 20 110 0]{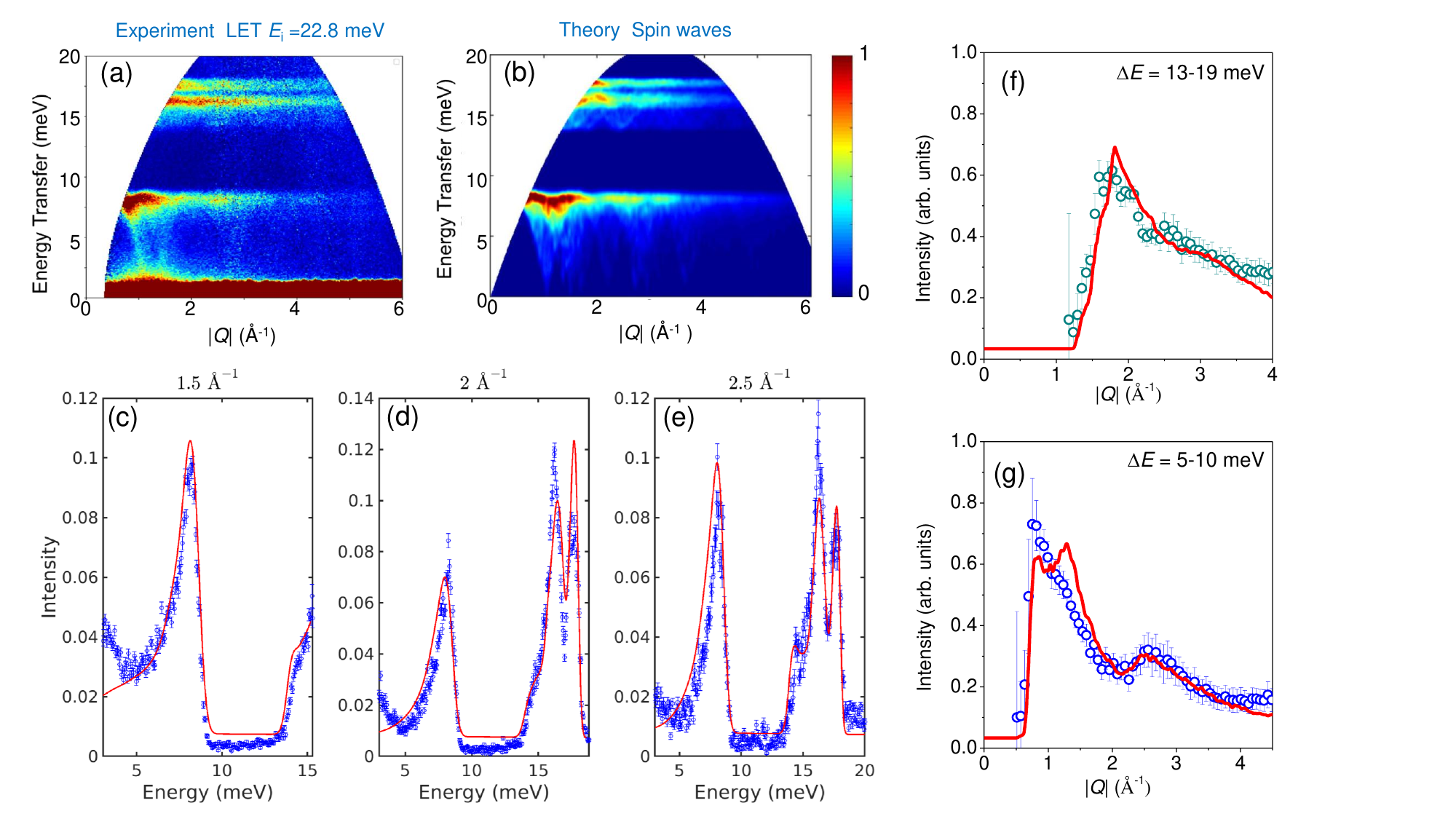}
	\caption{(Color online) (a) Color coded inelastic neutron scattering intensity maps, energy transfer versus momentum transfer ($Q$) at 5~K of PbMn$_{2}$Ni$_6$Te$_2$O$_{18}$ measured with an incident energy $E{_i}$=22.8~meV on LET. (b) The simulated spin-wave spectra using {\tt SpinW} program~\cite{Toth_2015} with the exchange parameters given in Table I. (c-e) Experimental (symbols) and simulated (solid line) $Q$-integrated Energy transfer versus Intensity plots at 1.5 \AA$^{-1}$, 2 \AA$^{-1}$~and 2.5 \AA$^{-1}$. (f-g) Experimental and simulated scattering intensity as a function of Q for the  energy transfer summed over 13 to 19~meV and 5 to 10~meV.}
	\label{FIG:insfit}
\end{figure*}	
 We have used linear spin wave theory for the analysis of the observed spin waves at 5 K. The analysis was carried out using the {\tt SpinW} program ~\cite{Toth_2015}. In the first stage of the analysis, we started with the six exchange parameters calculated using DFT (see Fig.~\ref{FIG:Unitcell} and given in Table-~\ref{I}). However, the calculated  energies of two bands of spin wave excitations were found to be near 13~meV, and 23~meV which were higher than the experimentally observed values of 8 and 18~meV, (see Fig.~\ref{FIG:SpinWSimu}, in the Appendix-D). DFT tends to predict larger exchange parameters than experimental values due to approximations in the exchange-correlation functional, as reflected in the overestimated spin wave excitation energies \cite{koo_crvo4,KOO2009602,grau2003cation}. In order to determine the values of the six exchange parameters, we fitted the full 2D spectrum $S(|Q|,\omega)$ at 5 K in the {\tt SpinW} program. The fitted value of the exchange parameters are given in Table-I. It is clear that the sign of the fitted exchange parameters are in agreement with the DFT exchange parameters as well as with the reported magnetic structure ~\cite{acs.inorgchem.pbmn2}. However, it is also clear that the DFT value of J$_1$ (and J$_6$) is a factor of 2 larger than that obtained from the fitting experimental data. A very similar behavior is observed for the DFT J$_2$, which is a factor of 2.5 larger than that of the fitted value from the experimental data. The DFT values of J$_3$, J$_4$ and J$_5$ are very close to the value obtained from the fitting. 
 
\begin{figure} 
	\includegraphics[width = \linewidth, trim= 60 90 70 50]{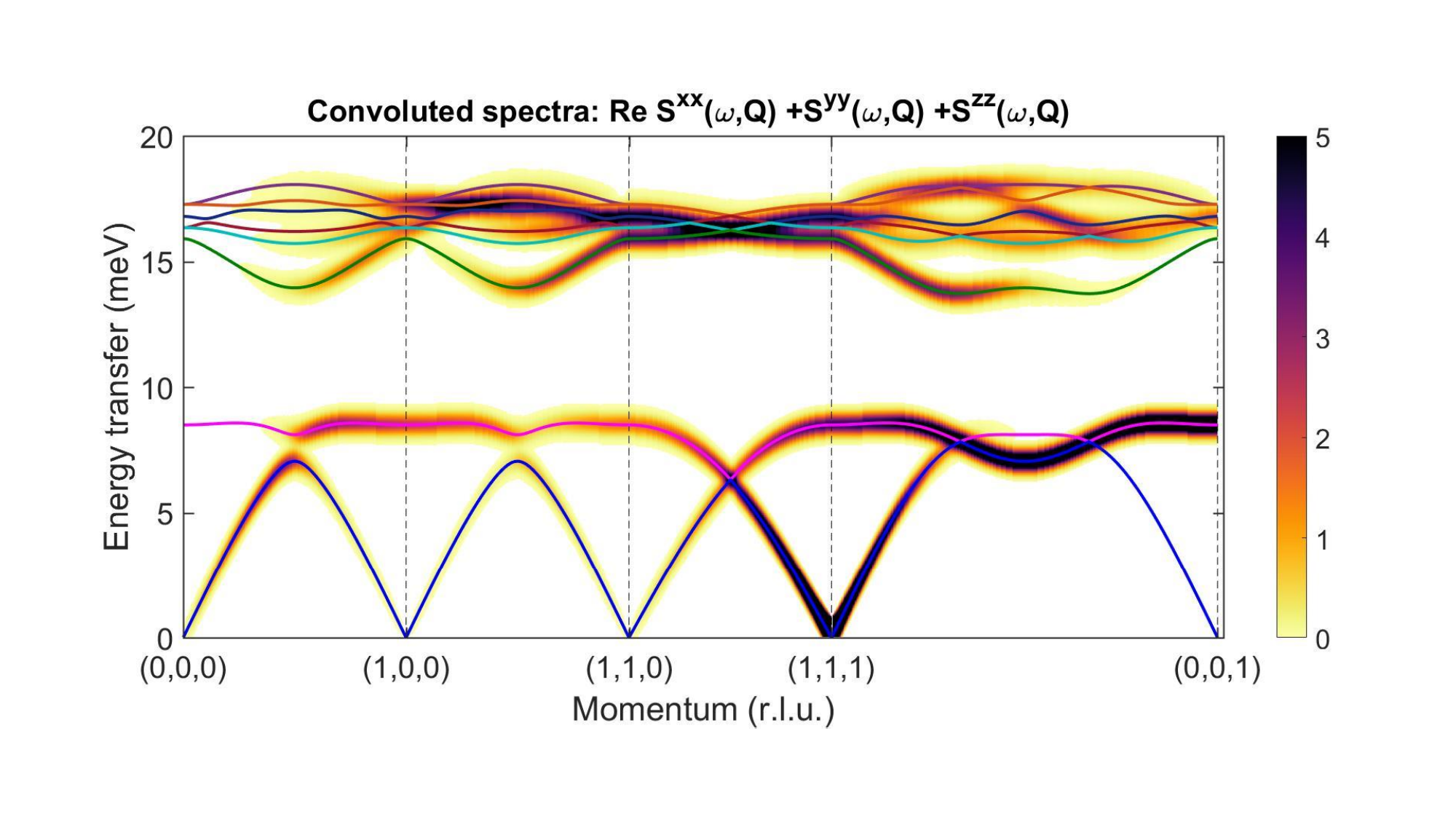}
	\caption{(Color online) The dispersion relations and the structure factors for the Hamiltonian of PbMn$_{2}$Ni$_6$Te$_2$O$_{18}$ obtained from INS fitting (Table I).} 
	\label{FIG:Dispersion}
\end{figure}

 The simulated spin-wave dispersions using the fitted values of the exchange parameters are given in Fig.~\ref{FIG:insfit}(b) and the experimental data are plotted in Fig.~\ref{FIG:insfit}(a), which reveal good agreement between the experimental and simulated results. In Fig.~\ref{FIG:insfit}(c)-(e), we have plotted the experimental Q-integrated energy versus intensity cuts at three Q-positions along with the fitted results (solid line). Further, Figs.~\ref{FIG:insfit}(f-g) show the experimental results and simulated results for one-dimensional energy integrated Q vs Intensity cuts at 13-19 meV and 5-10 meV, respectively.  These results also show good agreement between the experimental and theoretical results. The spin wave simulation with the optimized values of the exchange parameters captures the main features of the excitation spectrum (Figs.~\ref{FIG:insfit}(b-g)). The simulated dispersion relations and the structure factors, the latter convoluted with the instrumental energy resolution, are shown in Fig.~\ref{FIG:Dispersion} for the principal directions, which reveals eight spin wave modes arising from the Ni and Mn moments. The intensity of the spin wave modes is stronger for the (1~1~0) [Q = 1.354~$\AA$$^{-1}$] to (1~1~1) [Q = 1.531~$\AA$$^{-1}$] to (0~0~1) [Q = 0.713~$\AA$$^{-1}$] directions for the low energy branches and along (1~0~0) [Q = 0.782 ~$\AA$$^{-1}$] to (1~1~0) to (1~1~1) for the high energy branches. These results will be very useful in any future study of spin wave using the single crystals of PbMn$_2$Ni$_6$Te$_3$O$_{18}$.
 
 It should be noted that both our simulated spin wave and the experimental data exhibit no clear evidence of a spin gap. However, considering the instrumental resolution ($\sim 0.032$~meV for E$_i$ = 2.2 meV), we cannot exclude the possibility of a spin gap in the low-lying excitations. The gapless-like density of states was also reported from the finite value of the Sommerfeld specific heat coefficient ($\gamma$ =51.03(1) mJ~mol$^{-1}$~K$^{-2}$) for PbMn$_2$Ni$_6$Te$_3$O$_{18}$ ~\cite{PhysRevB.103.064401}. Given that the Haldane gap is related to the exchange parameters (E$_g$=0.4$\vert$J$\vert$), we should expect the gap value to be 1.05~meV (if we use J$_1$) or 1.68 meV (if we use J$_3$) in PbMn$_2$Ni$_6$Te$_3$O$_{18}$. However, considering the inter-chain interaction J$_3$ $>$ J$_1$ intra-chain interaction, which suggests  that this system is not
really a quasi-one-dimensional and supports the absence of a Haldane gap.  Further low energy inelastic neutron scattering measurements on a single crystal sample of PbMn$_2$Ni$_6$Te$_3$O$_{18}$ would be highly desired to shine light on the nature of the flat band observed near 1 meV.  Furthermore, it is interesting to compare the exchange parameter J=7.41 meV (nearest-neighbor, which is J$_1$ in our definition) which was estimated from the Raman scattering study, which is again a factor 2.83 larger than our estimated value from the fitting the spin waves ~\cite{PhysRevB.103.064401}.  

 Looking at the fitted exchange parameters in Table ~\ref{I}, it is clear that the strongest exchange parameter is J$_3$, between the Ni-chains (inter-chain), which is  at a distance of 3.758 $\AA$ with Ni-O$_1$-Ni bond angle of 123.763 $\degree$, where NiO$_6$ octahedra are connected through corner-sharing. The J$_1$ exchange is the second strongest. It connects spins with the shortest Ni-Ni distance of $2.943 \AA$ (with bond angles of 84.378 $\degree$ and 91.338 $\degree$), along the chain (intra-chain). There exist Ni-Ni structural dimers on this length scale, which are connected through the edge-sharing NiO$_6$ octahedra. The strength of J$_1$ can be attributed to the overlap between Ni-orbitals.  This is consistent with the electronic structure calculation, which reveals the Ni-d$_{x^2-y^2}$ contribution is dominant just below the Fermi level in the spin-up channel, while the Ni-d$_{yz/xz}$ orbitals exhibit significant presence in the spin-down channel. A very similar orbital effect on the exchange parameters has been observed in $\alpha$-Cu$_2$V$_2$O$_7$ where the third-nearest-neighbor exchange interaction is the strongest among the three exchange parameters between Cu dimers~\cite{PhysRevB.94.144426}. In the frustrated spin chain compound Li$_3$Cu$_2$SbO$_6$, density functional theoretical calculations and inelastic neutron scattering experiments show that the next-nearest neighbor (NNN) antiferromagnetic (AFM) exchange (J$_{AFM}$ = 31 meV) is stronger than the NN ferromagnetic
(FM) exchange (J$_{FM}$ = \textminus21 meV), indicating the importance of the orbital degrees of freedom~\cite{PhysRevB.103.174423}.


\section{Conclusion}

We have presented the theoretical DFT calculated exchange parameters and experimental results of inelastic neutron scattering on the quasi-one dimensional Ni-chain compound PbMn$_{2}$Ni$_6$Te$_2$O$_{18}$ to understand the nature of magnetic exchange interactions. Our inelastic neutron scattering study at 5 K on a powder sample reveals two bands of magnetic excitations, the first near 8 meV and the second near 18 meV originating from the antiferromagnetic zone center near Q=1 $\AA$ at 5 K. At 100 K (which is above T$_N$ = 86 K), the observed spin wave at low temperatures transforms into broad diffuse scattering indicating the presence of short range magnetic correlations due to magnetic frustration. We have analyzed the magnetic excitations based on the linear spin wave theory and compared the experimentally estimated exchange parameters with the DFT calculations. Although DFT gives good starting values of the exchange parameters, with correct sign (i.e., FM or AFM), the energy scale is higher than the experimental data. Our analysis reveals that the value of the exchange parameter at the largest distance (d=3.654 $\AA$) J$_3$=4.21(8) meV between Ni dimers is the strongest among the allowed six exchange parameters, which is due to the largest bond angle 123.673 $\degree$ between  Ni-O$_1$-Ni as well as the corner-sharing of NiO$_6$ octahedra. Our study will generate interest in the physics of ferrotoroidic PbMn$_{2}$Ni$_6$Te$_2$O$_{18}$ and further work using single crystal neutron scattering is highly desired. Furthermore, the results of the present study will form a starting point for analysis of spin wave excitations of many compounds of lead manganese nickel tellurium oxides family with a non-magnetic atom on the manganese site~\cite{sivakumar2020new}.

\section*{Acknowledgments}
 D. T. A. and S.U.  thank Engineering and Physical Sciences Research Council, UK for funding (Grant No. EP/W00562X/1) and D.T. A. tank the CAS for PIFI Fellowship. D. T. A. would like to thank the Royal Society of London for International Exchange funding between the United Kingdom and Japan and Newton Advanced Fellowship funding between the United Kingdom and China. A. P. and H. G. extend their gratitude to the Computer Division at RRCAT for providing the scientific computing facilities. Financial support for this study was provided to A. P. by HBNI-RRCAT. D.P. acknowledges the financial support by the Oxford-ShanghaiTech Collaboration Project and EPSRC [Grant No. EP/T028637/1]. Experiments at the ISIS Neutron and Muon Source were supported by beamtime allocations RB2190121 and RB2390101 from the Science and Technology Facilities Council. Data are available at  \href{https://doi.org/10.5286/ISIS.E.RB2390101} {https://doi.org/10.5286/ISIS.E.RB2390101} and 
 \href{https://doi.org/10.5286/ISIS.E.RB2190121-1} {https://doi.org/10.5286/ISIS.E.RB2190121-1}.
\\
\\
Data Availability Statements: The neutron data are available via reference \cite{RB2190121, RB2390101} and the rest of the data are available upon reasonable request from the authors.

\appendix 
\section{Optical Absorption Spectroscopy}

\textbf{Experimental Details:} The bandgap of synthesized samples was estimated using optical absorption spectroscopy measurements carried out in the diffuse reflectance mode (DRS). These measurements were conducted in the energy range 1.12 eV up to 6.50 eV. The temperature-dependent measurements were carried out using a self-developed high-temperature setup~\cite{gupta2022temperature, mishra2017electronic, gupta2022probing, sivakumar2020new}.

\begin{figure}[h]
    \centering
    \subfigure[]{
        \includegraphics[width=0.45\textwidth]{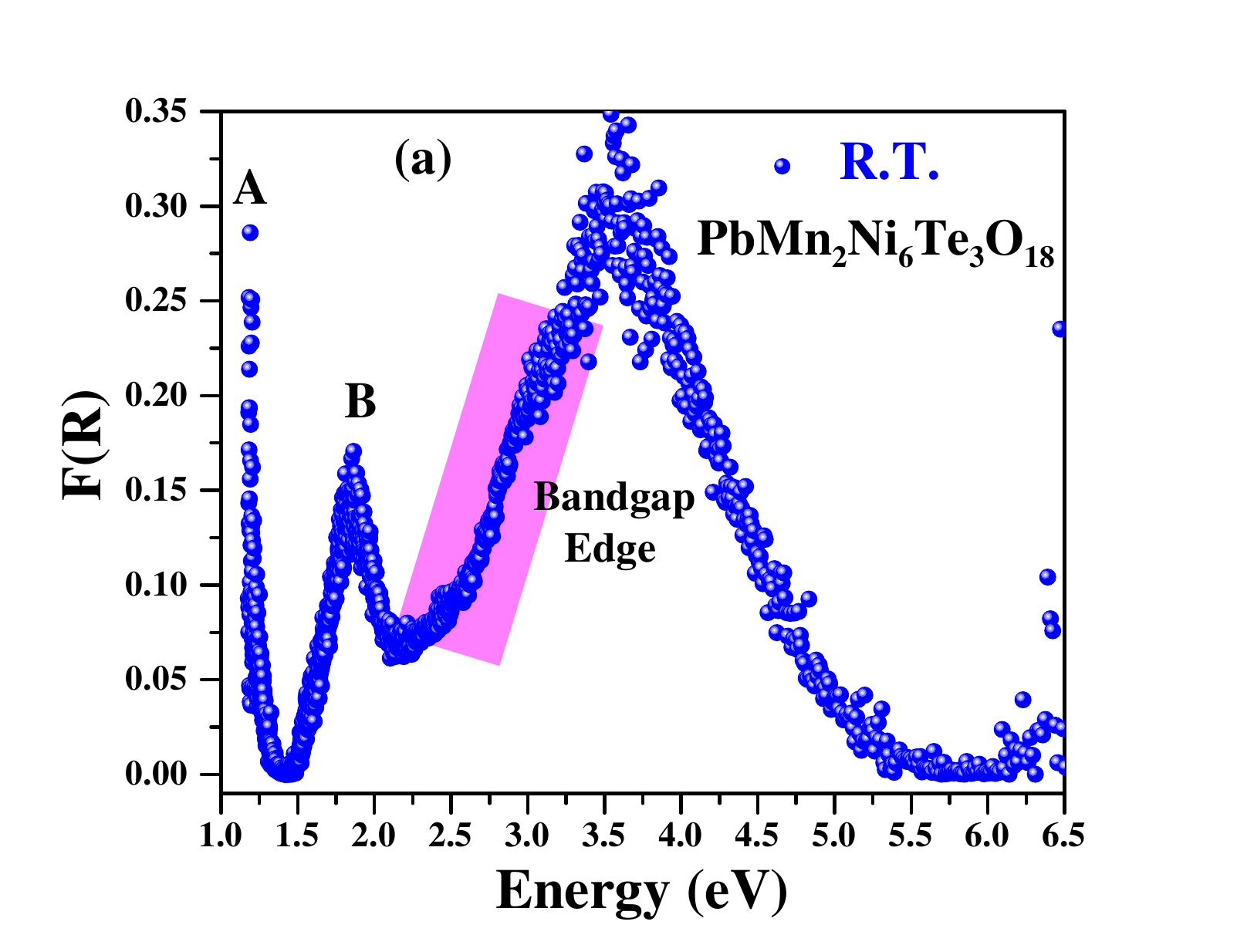} 
    }
    \hfill
    \subfigure[]{
        \includegraphics[width=0.45\textwidth]{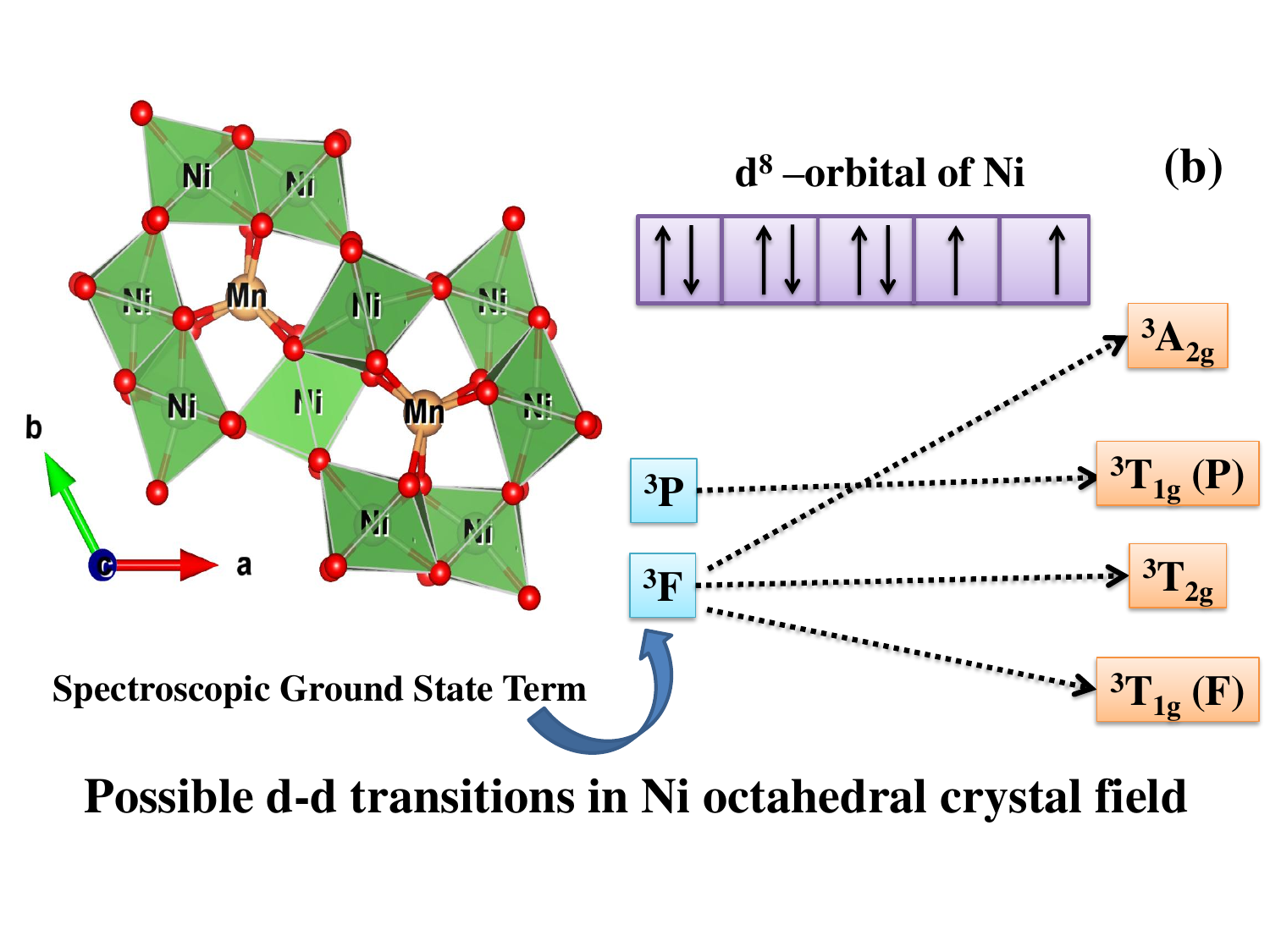} 
    }
    \caption{(a) Optical absorption spectra of PbMn$_2$Ni$_6$Te$_3$O$_{18}$ at room temperature (R.T.). (b) Possible signature of d-d transitions in the Ni octahedral crystal field.}
    \label{fig:absorption}
\end{figure}

In Figure~\ref{fig:absorption} (a), we show the optical absorption of PbMn$_2$Ni$_6$Te$_3$O$_{18}$ at room temperature. The features marked by ‘A’ and ‘B’, arise from the d-d transition of NiO$_6$ octahedra~\cite{czaja2018absorption}, and the region between 2.2 eV up to 3.0 eV highlighted in pink may correspond to absorption across the bandgap of PbMn$_2$Ni$_6$Te$_3$O$_{18}$. Density functional analysis can be useful in this regard. 

The possible d-d transitions in the Ni octahedral crystal field is depicted in Figure~\ref{fig:absorption} (b). Here, only the shoulder of the $d-d$ transition $^3A_{2g} \rightarrow ^3T_{2g}$ (peak ‘A’) in the infrared region has been probed due to instrument energy limitations. Further, two $d-d$ transitions $^3A_{2g} \rightarrow ^3T_{1g} (F)$ and $^3A_{2g} \rightarrow ^3T_{1g} (P)$ (peak ‘B’) have been observed in the visible region between the energy range 1.5–2.1 eV. The temperature-dependent DRS measurements have been carried out from room temperature to 485 K.

\begin{figure}[h]
    \centering
    \subfigure[]{
        \includegraphics[width=0.45\textwidth]{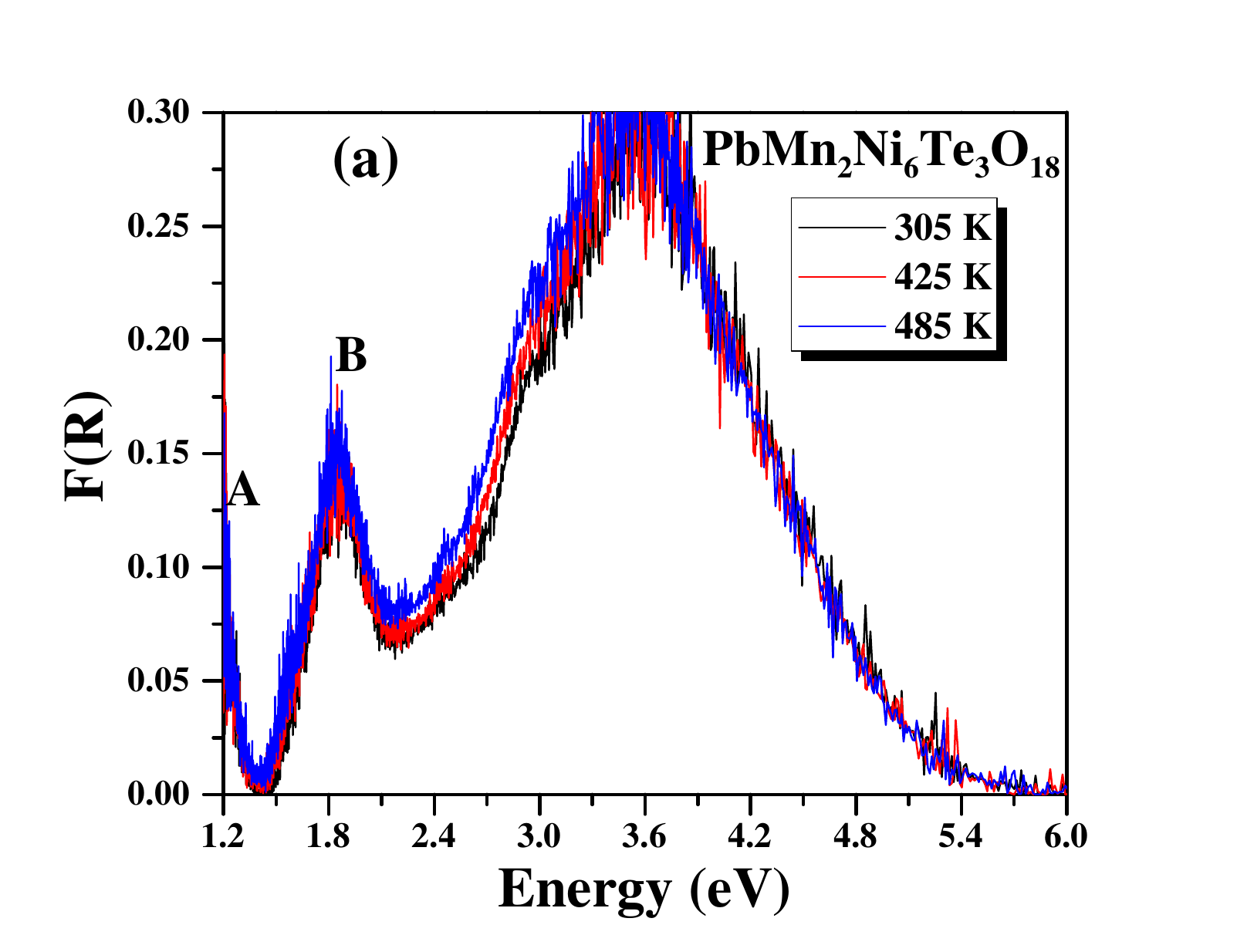} 
    }
    \hfill
    \subfigure[]{
        \includegraphics[width=0.45\textwidth]{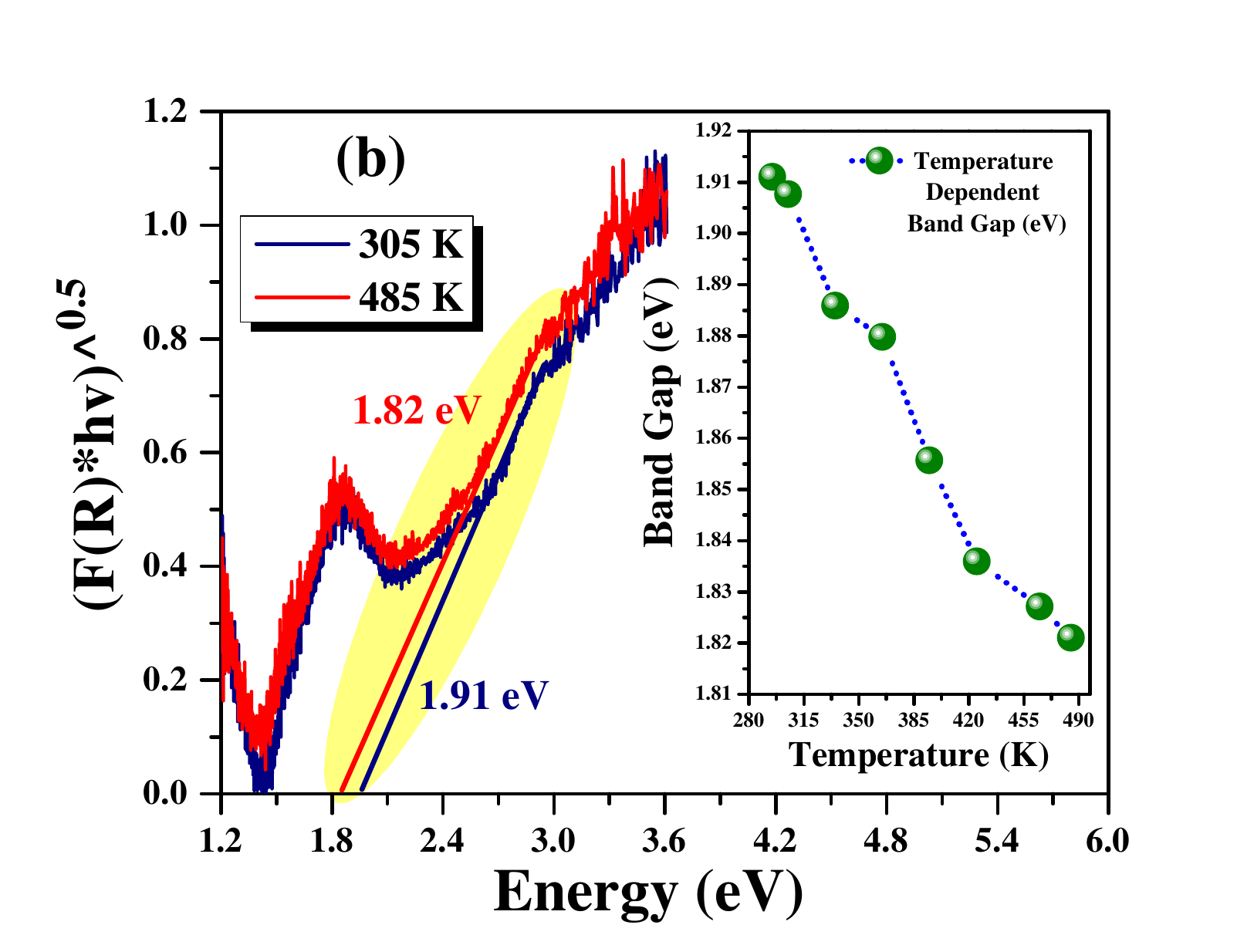} 
    }
    \caption{(a) Temperature-dependent optical absorption spectra of PbMn$_2$Ni$_6$Te$_3$O$_{18}$ at 305, 425, and 485 K. (b) Representative Tauc plot with a linear fit for PbMn$_2$Ni$_6$Te$_3$O$_{18}$ at 305 and 485 K. Inset: Temperature-dependent bandgap values for PbMn$_2$Ni$_6$Te$_3$O$_{18}$ calculated using the Tauc plot.}
    \label{fig:tauc}
\end{figure}

Figure~\ref{fig:tauc} (a) represents the temperature-dependent optical absorption spectra for PbMn$_2$Ni$_6$Te$_3$O$_{18}$ at 305, 425, and 485 K, respectively. A systematic red shift has been observed in peak ‘B’ of the visible region with increasing temperature, which may correspond to thermal expansion~\cite{gupta2022temperature}. The peak broadening of peak ‘B’ has also been seen with increasing temperature, which can be correlated with the distortion of NiO$_6$ octahedra. Due to the distortion in NiO$_6$ octahedra, some of the forbidden transitions become allowed, and hence peak broadening can be observed with increasing temperature~\cite{gupta2022temperature}.

\textbf{Bandgap Estimation:} To estimate the values of the bandgap, the diffuse reflectance spectra were converted into absorption using the Kubelka-Munk equation:

\begin{equation}
F(R) = \frac{K}{S} = \frac{(1 - R_{\infty})^2}{2R_{\infty}}
\end{equation}

\noindent where $K$ and $S$ are the Kubelka-Munk absorption and scattering functions, respectively, and

\begin{equation}
R_{\infty} = \frac{R_{sample}}{R_{reference}}.
\end{equation}

BaSO$_4$ powder was used as a reference. For diffuse reflectance of samples, it was assumed that reflected light scatters in a perfectly diffuse manner, and in such cases, the scattering function $S$ is nearly constant with wavelength. The Kubelka-Munk function, which is related to the absorption coefficient $(\alpha)$, can be written as:

\begin{equation}
F(R) \propto \alpha \propto \frac{(h\nu - E_g)^{1/n}}{h\nu}
\end{equation}

\begin{equation} \label{eq_4}
(\alpha h\nu)^n = A (h\nu - E_g)
\end{equation}

In equation~(\ref{eq_4}), ‘n’ has a value of 2 and 1/2 for direct and indirect bandgap transitions, respectively~\cite{gupta2022temperature}. We have considered PbMn$_2$Ni$_6$Te$_3$O$_{18}$ as an indirect bandgap ($n= 1/2$) material for room and high-temperature bandgap determination. The Tauc plot equation was used to obtain the bandgap of PbMn$_2$Ni$_6$Te$_3$O$_{18}$. The plot between $(\alpha h\nu)^n$ versus $h\nu$ yields a straight line, and the intercept/slope gives the bandgap value.

Figure~\ref{fig:tauc} (b) represents the Tauc plot with a linear fit at two different temperatures, 305 and 485 K, with bandgap values of 1.91 eV and 1.82 eV, respectively. The inset in Figure~\ref{fig:tauc} (b) shows the variation of bandgap with temperature. It was observed that with increasing temperature, the bandgap decreases in accordance with the Warshne relation.

\section{Inelastic measurements from MARI}
\begin{figure}
 \centering
	\includegraphics[width = \linewidth, trim=50 50 180 60, clip]{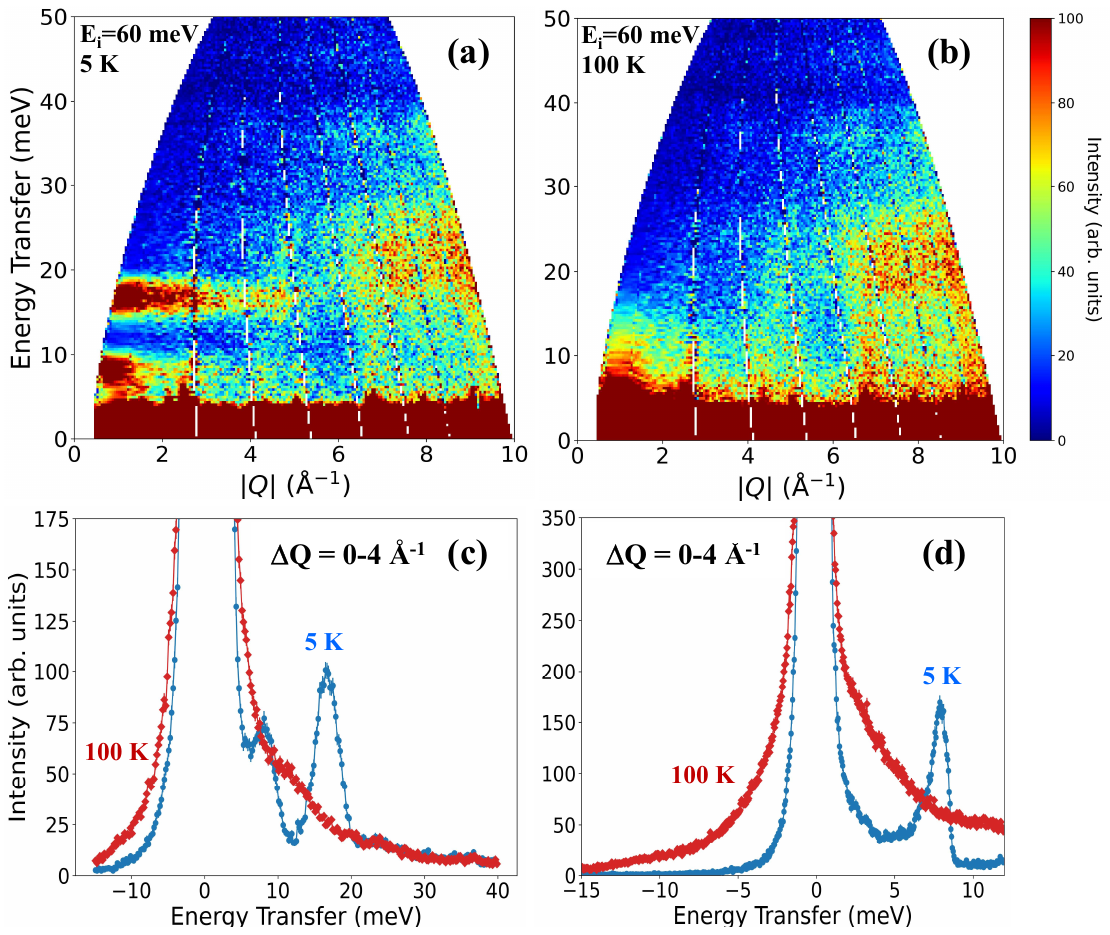}
	\caption{(Color online) Color coded intensity maps, Energy transfer versus Momentum transfer measured on MARI (a) at 5 K and (b) at 100 K (b) of PbMn$_{2}$Ni$_6$Te$_2$O$_{18}$.(c) and (d) show the Q-integrated (0 to 4 \AA$^{-1}$) Energy versus Intensity cuts at 5 K and 100 K, using E$_i$=60 and 17.7 meV, respectively.}
	\label{FIG:FigMARI}
\end{figure}

Here we present the results of our INS measurements from MARI at 5 K and 100 K with an incident energy of E$_i$ = 60 meV and 17.7 meV using the Gd-chopper frequency of 200 Hz obtained using the repetition rate multiplication (RRM) method, which gave both the energy data in the same run, see Fig. ~\ref{FIG:FigMARI}. As observed on LET, we have seen two bands of spin wave at 5 K (Fig.~\ref{FIG:FigMARI}(a, c) and diffuse scattering at 100 K(Fig.~\ref{FIG:FigMARI}(b, d) on MARI. Further, MARI data do not reveal any magnetic excitations above 20 meV at low-Q, which confirmed that the highest energy of the spin wave excitations is 18 meV.

\section{Low energy inelastic measurements from LET}
\begin{figure}
 \centering
	\includegraphics[width = \linewidth, trim=0 50 150 60, clip]{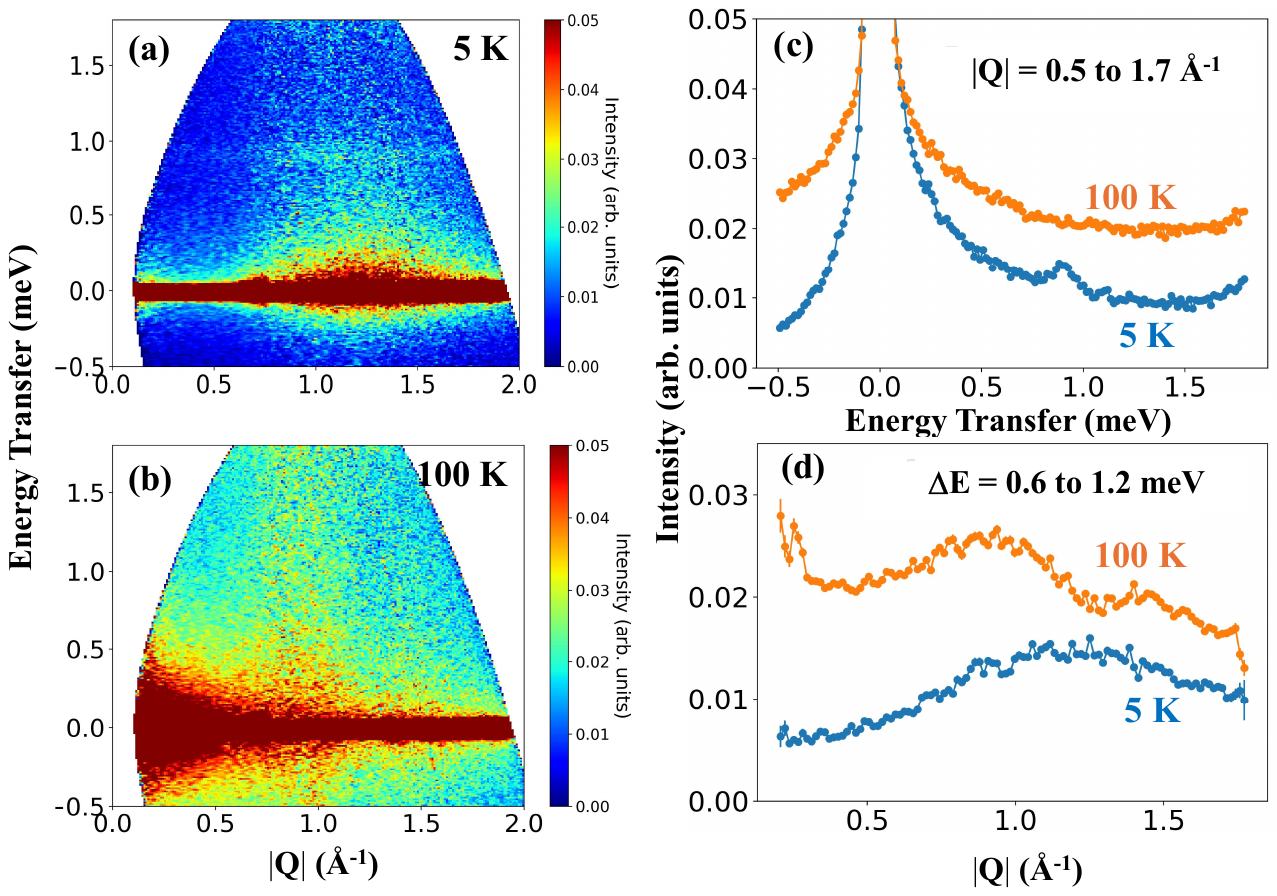}
	\caption{(Color online) Color coded intensity maps, Energy transfer versus Momentum transfer measured on LET  using E$_i$= 2.2 meV (a) at 5 K and (b) at 100 K (b) of PbMn$_{2}$Ni$_6$Te$_2$O$_{18}$.(c) shows the Q-integrated (0.5 to 1.7 \AA$^{-1}$) Energy versus Intensity cuts at 5 K and 100 K and (d) shows the Energy integrated (0.6 to 1.2 meV) Q versus Intensity cuts at 5 K and 100 K.}
	\label{FIG:Fig2p2meV}
\end{figure}

Here we present the results of our low energy INS measurements from LET at 5 K and 100 K with an incident energy of E$_i$ = 2.2 meV see Fig. ~\ref{FIG:Fig2p2meV}. We can see the gapless excitations withing the instrument resolution. We have also observed low energy inelastic peak near 0.92 meV (Fig.~\ref{FIG:Fig2p2meV}(a,c)) at 5 K, which was also observed in the data of E$_i$=3.7 meV. The low energy inelastic peak near 0.92 meV disappears at 100 K (Fig.~\ref{FIG:Fig2p2meV}(c)), confirming its magnetic origin. The energy integrated, momentum transfer (Q) versus Intensity plot at 5 K reveals a broad peak near 1.2~$\AA$$^{-1}$ (Fig.~\ref{FIG:Fig2p2meV}(d)), which may suggest its origin is related to magnetic frustration and not related to phonon modes as their intensity should increases with increasing Q as Q$^2$, which is not the case.

\section{Simulated Spin wave using DFT exchanges}
\begin{figure}
 \centering
	\includegraphics[width = \linewidth, trim=5 150 05 35, clip]{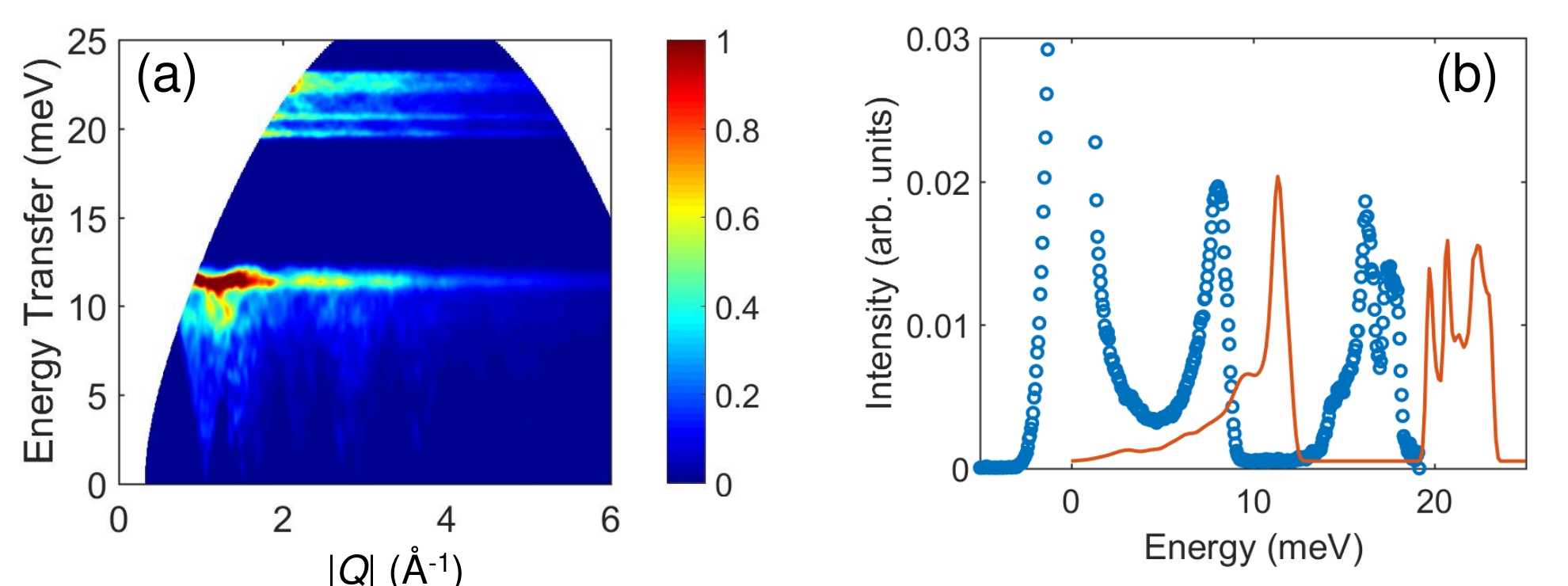}
	\caption{(Color online) The simulation of the spin waves using 
 DFT exchange parameters, (a) 2D color coded intensity maps, Energy transfer versus Momentum transfer and, (b) Q-integrated 1D cut, experimental data (blue circles) and simulated results solid line)  of PbMn$_{2}$Ni$_6$Te$_2$O$_{18}$}
	\label{FIG:SpinWSimu}
\end{figure}

Here we present the simulation of spin wave using the value of the exchange parameters obtained from the DFT calculation (see, Table~\ref{I}) to check whether these exchange parameters explain the observed spin wave excitations without any adjustment. Fig.~\ref{FIG:SpinWSimu}(a) shows the simulated spin wave excitations in 2D color plot, while  Fig.~\ref{FIG:SpinWSimu}(b) shows the comparison of  1D-cut from the data (blue circles) and simulated spectra (red solid line). Although the spin wave excitations obtained from the DFT exchange parameters have similarity with the experimental data, but the DFT exchange parameters give higher energies for the excitations compared with the data. This suggested that one needs to use least-square fitting method to obtained the exchange parameters. The value of the exchange parameters obtained from the fitting are given in Table~\ref{I}.

\section{Spin configurations}
Here we have presented the seven magnetic spin configurations used in the DFT  calculation and their corresponding energies are given in Eqs.(1-7). The exchange interaction network configurations between spins of Ni$^{2+}$ and Mn$^{2+}$  as well as their alignments that follow Heisenberg interaction obeyed by Eq (1-7) are presented in the Fig. 13. The AFM3 configuration is the ground state of PbMn$_2$Ni$_6$Te$_3$O$_{18}$.

\begin{figure*}
 \centering
	\includegraphics[width = \linewidth]{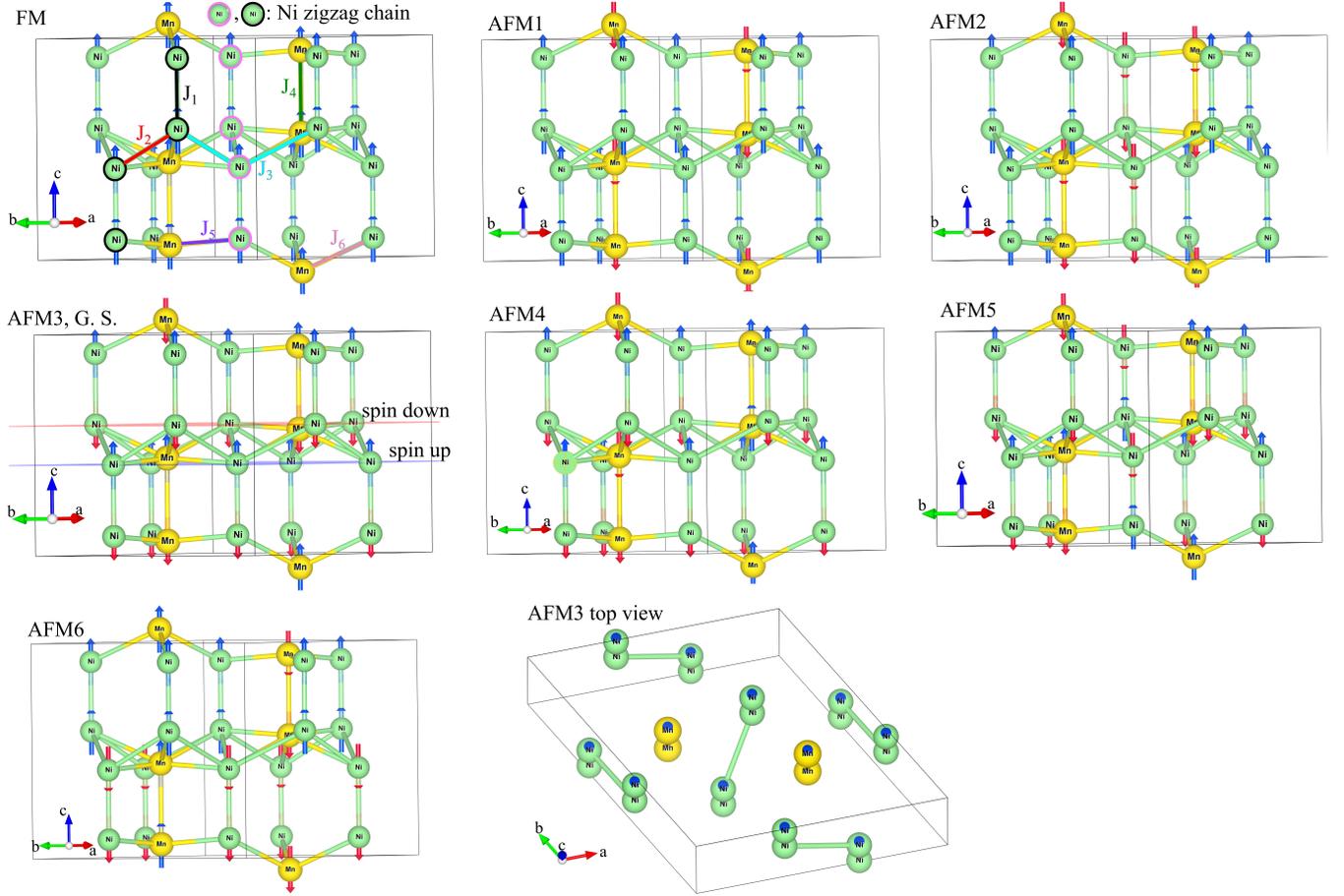}
	\caption{(Color online) All seven spin configurations used in the DFT calculations for PbMn$_{2}$Ni$_6$Te$_2$O$_{18}$, showing only the magnetic atoms. See equations (1 --- 7) for their corresponding energies. The exchange interactions are depicted, with the two Ni zigzag chains highlighted by black and magenta circles (see top Left FM configuration).
 }
	\label{FIG:Spin_config}
\end{figure*}

\clearpage
\clearpage
	\bibliography{library}

\end{document}